\documentclass[a4paper,11pt]{article}
\usepackage{siunitx}
\usepackage{jheppub}
\usepackage{array}
\usepackage{float}
\usepackage{booktabs}
\usepackage{makecell}
\usepackage{caption}
\usepackage{amsmath}
\usepackage{graphicx,tabularx}

\usepackage{gensymb}
\usepackage{url}
\usepackage{footmisc}
\usepackage{amsfonts}
\usepackage{cancel}
\usepackage{color}
\usepackage{multirow} 
\usepackage{amssymb}
\usepackage{pifont}
\usepackage{epstopdf}
\usepackage{slashed}
\usepackage{comment}
\usepackage{booktabs} 
\usepackage{subfigure}
\usepackage{mathrsfs}
\usepackage{subcaption}
\usepackage[toc,page]{appendix}
\usepackage{mathtools}
\usepackage{romannum}
\usepackage[normalem]{ulem}
\usepackage{bbold}
\usepackage{xcolor}
\usepackage{tikz-feynman}
\usepackage{tikz}
\usepackage{placeins}
\usepackage{hyperref}
\usepackage{orcidlink}
\allowdisplaybreaks

\usepackage{accents}

\def\beq{\begin{equation}}
\def\eeq{\end{equation}}
\title{Drell-Yan Production of New Particles 
at Fixed-Target Experiments: Heavy Neutral Lepton as a Case Study
}
\author[]{Francis M. Burk$^1$\orcidlink{0009-0001-7398-8211},}
\affiliation[1]{Pittsburgh Particle Physics, Astrophysics and Cosmology Center, \\
Department of Physics and Astronomy, University of Pittsburgh, Pittsburgh, PA 15260, USA}
\emailAdd{frb25@pitt.edu}
\author[]{P. S. Bhupal Dev$^2$\orcidlink{0000-0003-4655-2866},}
\affiliation[2]{Department of Physics and McDonnell Center for the Space Sciences, \\
Washington University, St. Louis, MO 63130, USA}
\emailAdd{bdev@wustl.edu}
\author[]{Bhaskar Dutta$^3$\orcidlink{0000-0002-0192-8885},}
\emailAdd{dutta@tamu.edu}
\affiliation[3]{Mitchell Institute for Fundamental Physics and Astronomy, \\
Department of Physics and Astronomy, Texas A\&M University, College Station, TX 77843, USA}
\author[]{Tao Han$^1$\orcidlink{0000-0002-5543-0716},}
\emailAdd{than@pitt.edu}
\author[]{Aparajitha Karthikeyan$^3$\orcidlink{0000-0002-6934-1239},}
\emailAdd{aparajitha\_96@tamu.edu}
\author[]{Doojin Kim$^4$\orcidlink{0000-0002-4186-4265}}
\emailAdd{doojin.kim@usd.edu}
\affiliation[4]{Department of Physics, University of South Dakota, Vermillion, SD 57069, USA}

\preprint{
\begin{flushright}
PITT-PACC-2510 \\
MI-HET-877
\end{flushright}
}

\abstract{We demonstrate the sensitivity of Drell-Yan production processes from deep inelastic scattering in searches for beyond-the-Standard Model (BSM) physics at fixed-target or beam-bump experiments. We take heavy neutral leptons (HNLs) as a case study, produced from the decay of a light vector boson mediator with mass in the range of $2-20$ GeV, which itself is generated via the Drell-Yan process. The produced HNLs subsequently decay into Standard Model final states. We consider several current and future experiments, including SBND, DarkQuest, DUNE Near Detector (ND), and SHiP. Utilizing $\nu\pi^0$ and $\nu e^+e^-$ final states from HNL decays, we find that the Drell-Yan mechanism provides important contributions and significantly enhances the HNL search sensitivity, owing to the production of energetic final-state particles that are more readily detectable over the expected backgrounds. We find that at $90\%$ C.L.~sensitivity, for gauge couplings $g_{X} \sim 10^{-2}\ (10^{-3})$ and kinematically accessible mass range,  
SBND and DarkQuest can probe the HNL flavor mixing $|U_{\ell}| \sim 3\times 10^{-4}\ (10^{-3})$, whereas DUNE ND and SHiP  may extend the sensitivity down to the Type-I Seesaw prediction of $|U_{\ell}| \sim 10^{-5}$.  Finally, for our chosen benchmark $|U_{\ell}| = 10^{-3}$ outside of the current  experimental constraints, with a fixed mass ratio $m_{Z'}/m_N = 2.1$, and working within the $U(1)_{B-L}$, $U(1)_{B-3L_{\tau}}$, and $U(1)_{B}$ parameter spaces, we find that both SBND and DarkQuest can probe $g_{X} \sim 10^{-3}$, DUNE ND can reach $g_{X} \sim 10^{-4}$, and SHiP can probe down to $g_{X}\sim 5\times 10^{-6}$.  Our approach provides a powerful new technique to study HNL production at future fixed-target experiments and can readily be extended to other light BSM particle production within a broader class of dark sector models.}

\begin{document}

\titlepage
\maketitle

\section{Introduction}

The existence of new physics beyond the Standard Model (BSM) is well-motivated as a means to address several outstanding puzzles in contemporary particle physics, such as the origin of neutrino masses and mixing, the nature of dark matter, and the matter-antimatter asymmetry in the Universe. A wide range of experimental efforts is ongoing and planned to probe the scale of new physics. Among them, proton beam-based neutrino experiments provide a unique opportunity to explore light BSM particles with masses $\lesssim \mathcal{O}(10)$ GeV using high-intensity proton beams at facilities such as SBND~\cite{MicroBooNE:2015bmn}, ICARUS~\cite{ICARUS:2023gpo}, and DUNE~\cite{DUNE:2021tad}. When proton beams strike a target, they produce a high-intensity flux of $\gamma$, $e^\pm$, charged and neutral mesons, neutrons, etc., which can be utilized to probe dark sector states~\cite{Batell:2022xau}. The development of the Liquid Argon Time Project Chamber (LArTPC) technology, with its excellent energy and angular resolution, marks a significant advancement in neutrino fixed-target experiments by enabling precise particle identification~\cite{MicroBooNE:2020hho}. Additionally, ongoing and upcoming LArTPC detectors, such as SBND and DUNE Near Detector (ND), have cutting-edge technology such as \(\mathcal{O}(\rm{ns})\) timing resolution~\cite{MicroBooNE:2023ldj}, PRISM technology for spatial resolution~\cite{nuPRISM:2014mzw, DUNE:2021tad}, and more. This technology can be leveraged not only to increase their sensitivity reach but also to distinguish various BSM scenarios. Likewise, fixed-target experiments with higher-energy beams and shorter baselines, such as DarkQuest~\cite{Apyan:2022tsd} and SHiP~\cite{SHiP:2021nfo}, employ high-performance calorimetry, enabling access to complementary regions of parameter space and providing additional discriminating power. 

Many dark sector models introduce light scalar, pseudoscalar, vector, or axial-vector mediators 
in addition to sub-GeV dark matter candidates. Examples include an extension in the gauge sector to  $U(1)_{B-L}$, $U(1)_{L_\mu-L_\tau}$, $U(1)_{T3R}$, $U(1)_{A'}$, or in the scalar sector of Higgs-extended models, serving as potential portals to the dark sector~\cite{ 
Gori:2022vri}. Such mediators can be produced at proton beam facilities through processes such as charged and neutral meson decays, proton bremsstrahlung, Primakoff-like production, Compton-like processes, associated production, resonances, and electron bremsstrahlung. Once produced, the mediators can decay into BSM particles that couple to them, such as dark matter or heavy neutral leptons (HNLs) motivated from the seesaw mechanism. In addition, they can interact via scattering processes such as inverse Compton scattering, inverse Primakoff-like process, and Bethe-Heitler-like mechanisms.

At higher energies, the nucleons will undergo deep inelastic scattering (DIS). As such, the quark parton scattering becomes an important contribution. In this work, we focus on the Drell-Yan production processes via quark and anti-quark annihilation~\cite{Drell:1970wh} to probe light new physics at fixed-target experiments that have been mostly overlooked in the literature.
An important feature of Drell-Yan resonant production is that the final state particles produced in BSM scenarios tend to have significantly higher kinetic energy compared to the production from meson decays or processes involving electromagnetic fluxes, thus leading to efficient signal separation from the backgrounds. 
As a case study, we explore the search sensitivity for HNLs (henceforth denoted by $N$) produced via a light vector mediator (henceforth denoted by $Z'$). 
We will use parton distribution functions (PDFs) to estimate the cross sections and consider scenarios where the mediators decay primarily into HNLs. 
 We analyze the energy distributions of the HNL decay products and demonstrate the regions of the HNL parameter space that can be probed through Drell-Yan production, highlighting how this channel can be distinguished from other production mechanisms. Our approach is equally applicable to the searches for other BSM physics in the light dark sector.

The rest of this work is organized as follows. We introduce the theoretical framework featuring a new vector mediator $Z'$ and HNL $N$ in Section \ref{Sec:Model}. We briefly discuss some key features of the benchmark experiments that we consider in our study in Section~\ref{sec:Benchmark}. Section \ref{sec:signal} describes the generation of the signal relevant for fixed-target experiments. Section \ref{sec:backgrounds} details how we compare signal and background events for each experiment. In Section \ref{Sensitivity}, we present the projected sensitivities of these experiments for our process. We summarize our results and conclude in Section \ref{sec:summary}. The formulas of the HNL decay rates are compiled in Appendix~\ref{app: HNL Decays}.

%%%%%%%%%%%%%%%%%%%%%%%%%%%%%%%%%
\section{Theoretical Model Setting}
\label{Sec:Model}

We consider a simple $U(1)_X$ extension to the SM, which, in general, adds a new massive vector boson $Z'$ and HNL $N$. The new charge is parametrized by the couplings $g_X, g_{V,f}, g_{A,f}, g_{V,N}$ and $g_{A,N}$. The effective Lagrangian is expressed as
\begin{equation}
    \mathcal{L}_X \supset -\frac{1}{4}F'_{\mu\nu}F'^{\mu\nu} + \frac{1}{2}m_{Z'}^2Z_{\mu}'Z'^{\mu} +g_{X}Z_{\mu}'\Bigg(\sum_f\bar{f}\gamma^{\mu}(g_{V,f} + g_{A,f}\gamma_5)f +\bar{N}\gamma^{\mu} (g_{V,N} + g_{A,N}\gamma_5) N\Bigg),
\end{equation}
where $f$ is summed over all SM quarks and leptons $f=q,l$. There are many incarnations of the $U(1)_X$ gauge theory, depending on how the SM particles are charged under the new gauge group. In $U(1)_{B-L}$, the $Z'$ couples to SM fermions of all chiralities. The anomaly arising from this gauge symmetry may be removed by introducing three massive HNLs for each flavor. In the $U(1)_B$ leptophobic model, the gauge boson couples to baryons of all chiralities, but not to SM leptons. Although the SM leptons are uncharged, HNLs may still be introduced to cancel the mixed gauge-gravity anomaly $[R]^2[U(1)_{B}]$. The $[U(1)_{B}]^3$ anomaly can be removed by including a Peccei-Quinn term in the Lagrangian, and the remaining non-vanishing mixed anomalies can be removed by adding a generalized Chern-Simons term~\cite{Arun:2022ecj}. An anomaly-free $U(1)_B$ model can be constructed involving $N$ and new vector-like fermions (under the SM gauge group)~\cite{Carena:2004xs}. Alternatively, one can consider the $U(1)_{B-3L_i}$ class of models involving $N$, where the individual lepton flavors are isolated~\cite{Bauer:2020itv}. These models take a general form $U(1)_{B-xL}$, where $x=0$ corresponds to $U(1)_B$, $x = 1$ to $U(1)_{B-L}$ and $x_i = 3$ corresponds to $U(1)_{B-3L_i}$. As another example of models involving $N$, the $U(1)_{T_{3R}}$ couples only to right chiral fermions, proportional to its $T_3$ isospin. While a generic $U(1)_{T_{3R}}$ couples to all right-handed fermions, multiple variations of this model exist where the gauge boson can couple to selected generations~\cite{Dutta:2019fxn}. In this work, we present the parameter space of the $U(1)_{B-L}$, $U(1)_{B-3L_i}$, and $U(1)_{B}$ (involving $N$) models, as these scenarios permit different allowed ranges of couplings subject to existing experimental constraints. The fermion couplings for these specific model scenarios are summarized in Table \ref{ModelCouplings}. 
\begin{table}[tb]
\centering
\setlength{\tabcolsep}{1.5pt}  
\renewcommand{\arraystretch}{1.3} 
\footnotesize
\begin{tabularx}{0.5\linewidth}{X || c c ||c c || c c }
\toprule
  ~~~ & $g_{V,q}$ & ~~~$g_{A,q}$ & $g_{V,l}$ & ~~~$g_{A,l}$& $g_{V,N}$ & ~~~$g_{A,N}$ \\
\midrule
$U(1)_{B-L}$ & 1/3  & 0 & $-1$ & 0 & 0 & $-1$ \\
$U(1)_{B-3L_{\tau}}$  & 1/3  & 0 & $-3$ & 0 & 0 & $-1$ \\
$U(1)_{B}$  & 1/3  & 0 & 0 & 0 & 0 & $-1$ \\ 
\bottomrule
\end{tabularx}
\caption{Vector and axial-vector couplings of quarks, leptons and HNL for benchmark $U(1)$ models. Note that the HNL couplings to $Z'$ are purely axial-vector type because of the Majorana nature of HNLs assumed here. }
\label{ModelCouplings}
\end{table}
\indent When kinematically accessible, the $Z'$ can decay to the SM charged leptons $l^+l^-$, hadrons via $q\bar q$, as well as $\nu \bar \nu$ and $N N$. In general, the partial decay rate for $Z'\to f\bar{f}$ where $f$ can be a Dirac or Majorana fermion is given by
\begin{equation}
\Gamma(Z'\to f\bar{f}) = \frac{N_c}{S}\frac{g_X^2m_{Z'}\beta_f}{12\pi}\Bigg[(g_{V,f}^2 + g_{A,f}^2)\beta_f^2 +  g_{V,f}^2\frac{6m_f^2}{m_{Z'}^2}\Bigg],
\label{eq:width}
\end{equation}
where $S$ is the symmetry factor and $N_c$ is the color factor.
For example, the partial decay rates for these channels in the $U(1)_{B-L}$ model with the couplings provided in Table \ref{ModelCouplings} are 
\begin{align}
\Gamma(Z' \to l^+l^-) & =\frac{g_{B-L}^2m_{Z'}}{12\pi}  F_l, \quad 
{\rm with}~F_l = \beta_l \left(1 + \frac{2m_l^2}{m^2_{Z'}}\right), 
\label{eq:Zdec}
\\
\Gamma(Z'\to NN) &= \frac{g_{B-L}^2m_{Z'}}{24\pi} \beta_N^3,
\\ 
\Gamma(Z' \to \text{hadrons}) &= {\sum_q \Gamma(Z'\to q\bar{q})\over {\sum_q Q_q^2 F_q}} R(m_{Z'}) ,
\label{eq:Zdecay}
\end{align} 
where $\beta_f = (1-4m_f^2/m^2_{Z'})^{1/2}$ is the fermion speed in the center-of-mass frame. We note the different threshold behaviors between a Dirac fermion ($\beta_l$) and a Majorana fermion ($\beta^3_N)$. 
In Eq.~(\ref{eq:Zdecay}), $F_q$ is the kinematic factor in Eq.~(\ref{eq:Zdec}) with $l$ replaced by $q$ of electric charge $Q_q$ and the standard $R$-value parametrizes the hadronic contributions and can be evaluated following Refs.~\cite{Ilten:2018crw, Bauer:2018onh, ParticleDataGroup:2024cfk}, taking into account the Vector Meson Dominance approximation near resonances in 
the regime $m_{Z'}<1.65~\rm{GeV}$. 

\begin{figure}[tb]
    \centering
    \includegraphics[width=0.45\textwidth]{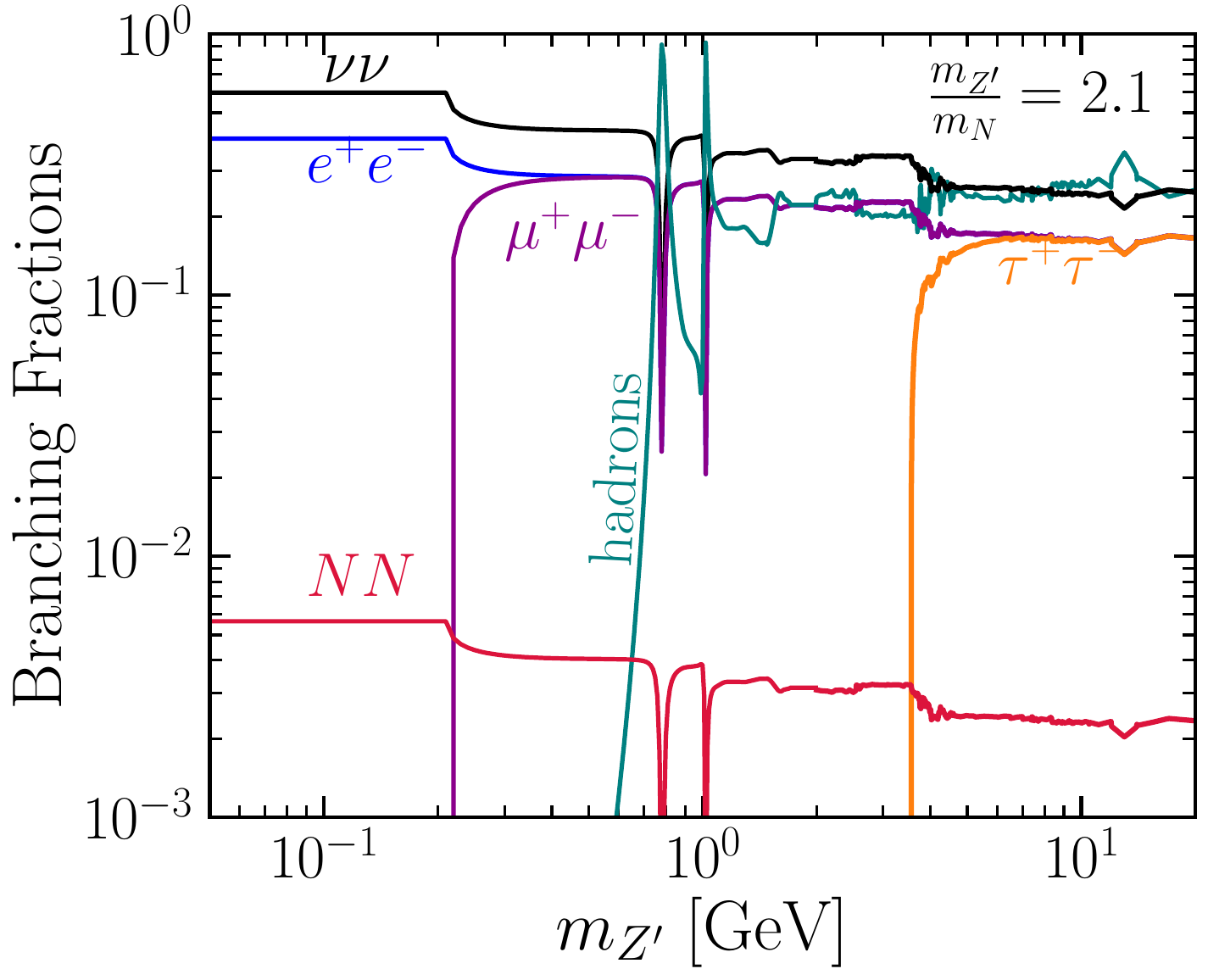}   
    \includegraphics[width=0.52\textwidth]{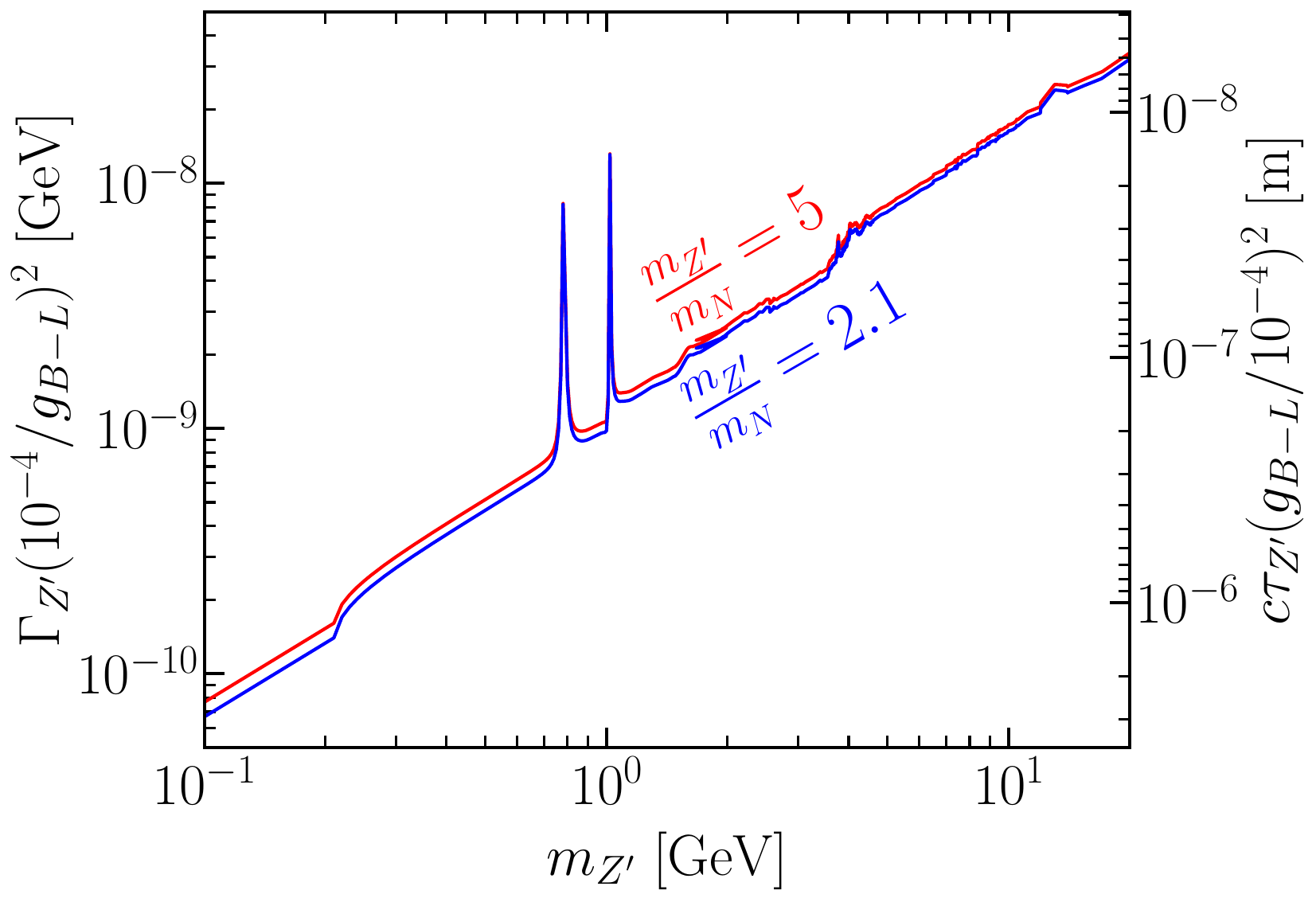}   
    \caption{The decay of $Z'$ in the $U(1)_{B-L}$ model versus its mass branching fractions (left panel), and total decay width and decay length (right panel). The sharp peaks near $m_{Z'} = 0.75$~GeV and $1$~GeV are due to the $\rho/\omega$, and $\phi$ meson resonances, respectively. In the left panel, we have fixed $m_{Z'}/m_N=2.1$, whereas in the right panel, we also show the $m_{Z'}/m_N=5$ case.} 
    \label{fig:GZp}
\end{figure}

The branching fractions and the total decay rate for the $Z'$ are presented in Fig.~\ref{fig:GZp}. In the region shown with $m_{Z'}<20~\rm{GeV}$, the $Z'$ decay rate is dominated by decays to  $l^+l^-$, $\nu\bar{\nu}$, and hadrons. The resonant structure is due to the hadronic resonances in the $R$-value measurement ~\cite{ParticleDataGroup:2024cfk}. Although the $NN$ mode is sub-dominant, the long-lived nature of HNLs makes them best suited for our study. This is the focus of the current work. Note that, depending on the $U(1)_X$ charges, the $Z'\to NN$ decay mode can also be the dominant one, which has been exploited in the context of high-energy collider searches~\cite{Das:2017flq, Das:2019fee}.

Regardless of the model choice, in this work we focus on $Z'$ arising from perturbative QCD production. Near light vector masses $m_{Z'} \sim 1$ GeV, particular care is required~\cite{Ilten:2018crw}. For masses $m_{Z'} \lesssim 1$ GeV, the production of the $Z'$ cannot be treated in the perturbative regime, and would require hadronic models. Therefore, we focus on the conservative $Z'$ mass range of $ m_{Z'} \geq 2 \text{ GeV}$ to remain within the perturbative QCD regime.  

Over the years, there have been many searches for additional $Z'$ vector bosons or $A'$ dark photon arising from a new $U(1)$ gauge group. Relevant to the $U(1)_{B-L}$ model, the work of the BaBar collaboration studying invisible states produced in a $e^+e^-$ collision~\cite{BaBar:2017tiz} and the LHCb search for dark photon decays $A'\to \mu^+\mu^-$~\cite{LHCb:2017trq} place strong upper bounds on a light $Z'$ coupling. Additional constraints on the $Z'$ mass-coupling space arise from old searches at SPEAR, DORIS, and PETRA~\cite{Carlson:1986cu, Ilten:2018crw}. For a $U(1)_{B-3L_{\tau}}$ model, the dominant constraint comes from Non-Standard Interaction-induced neutrino oscillations~\cite{Heeck:2018nzc, Coloma:2020gfv}. In the case of a $U(1)_B$ model, the lack of visible charged leptons in the final state results in a less constrained model parameter phase space. In our region of interest, the dominant constraints come from $\Upsilon$ decays~\cite{ ARGUS:1986nzm, Aranda:1998fr}, longitudinal enhancements~\cite{Dror:2017ehi, Dror:2017nsg} in $B_{u,d} \to KX$~\cite{Belle:2017oht} and $Z\to X\gamma$~\cite{DELPHI:1996qcc, L3:1997exg}, and from the lack of observed anomaly-canceling fermions~\cite{Dror:2017nsg, Dror:2017ehi, Dobrescu:2014fca, Ilten:2018crw}.    
With the $Z'$ search constraints in mind, we would like to pay particular attention to the $\tau$ flavor.  Consistent with the model-dependent bounds, we fix the following benchmark gauge couplings for each of our models for the rest of the presentation:
\begin{equation}
    g_{B-L}^{} = 10^{-4},\qquad 
    g_{B-3L_\tau}^{} = 10^{-3},\qquad 
    g_{B}^{} = 10^{-2}. 
    \label{eq:coup}
\end{equation}

\begin{figure}[tb]
    \centering    
    \includegraphics[width=0.98\linewidth]{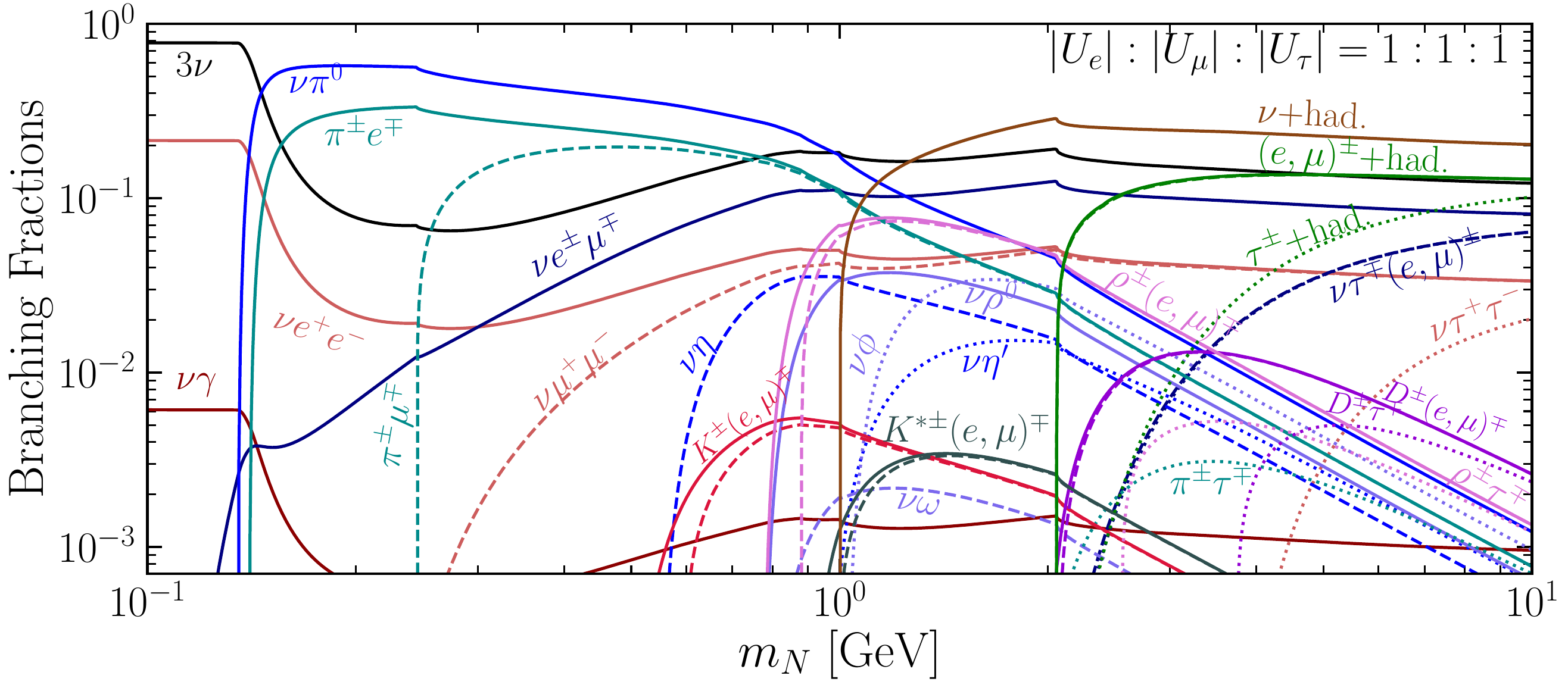}
    \includegraphics[width=0.98\linewidth]{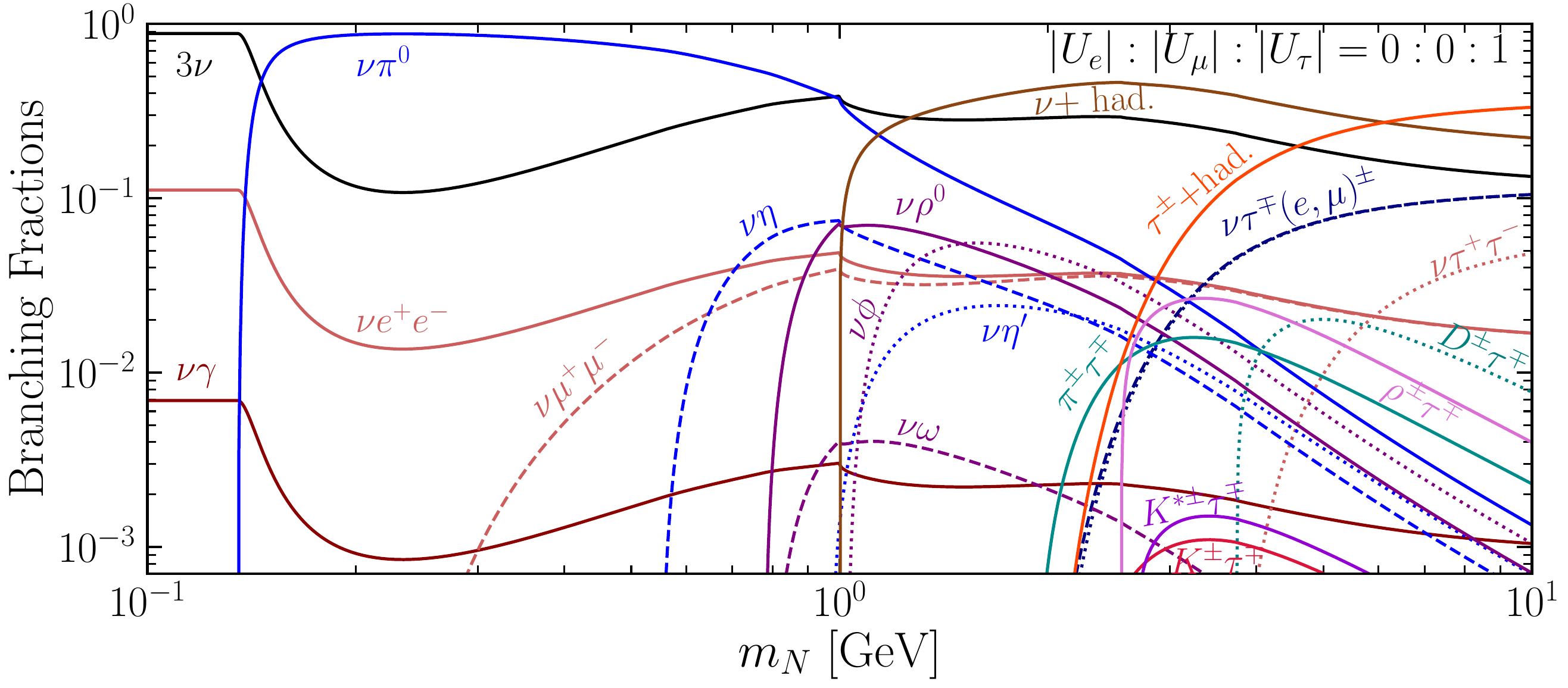}
    \caption{HNL decay branching fractions to various SM final states with (a) equal mixing angles with the three
    lepton generations (top panel), and (b) only $U_{\tau}$ mixing (bottom panel). 
    }
    \label{fig:BR_UT_1}
\end{figure}

In addition to direct gauge coupling of the $Z'$-$N$, the SM-singlet HNL can mix with one or more of the SM neutrino species through mass diagonalization. This is done by extending the leptonic PMNS mixing matrix by adding a new heavy state $N$. It can be either a (pseudo)Dirac or Majorana fermion, depending on its ultraviolet (UV)-completion. For our phenomenological study, we take $N$ to be Majorana in nature, without specifying any particular UV-complete theoretical model. 
The flavor eigenstates are written as
\begin{equation}
    \nu_\alpha = \sum_{i=1}^3 U_{\alpha i}\nu_i~+~U_{\alpha N}N \, .
\end{equation}
From now on, we use the shorthand notation for the mixing as $U_{\alpha} \equiv U_{\alpha N}$, where $\alpha$ specifies the lepton flavor $e,\mu,\tau$.
Through the mixing, the HNLs can decay to final state SM particles accessible by particle detectors. These decay rates of the HNLs to SM fields have been well-studied in the literature~\cite{Bondarenko:2018ptm, Ballett:2019bgd, Coloma:2020lgy, Capozzi:2024pmh}. 
The branching fractions of the HNL to SM final states are shown in Fig. \ref{fig:BR_UT_1} for two cases: (a) assuming equal mixing to all three generations of neutrinos, and (b) assuming mixing only with the $\tau$ flavor. A single HNL mixing with either a single neutrino flavor or multiple flavors is a common assumption made in the phenomenological HNL studies (see e.g.,~Refs.~\cite{Atre:2009rg, Deppisch:2015qwa, deGouvea:2015euy, Cai:2017mow, Bolton:2019pcu, Abdullahi:2022jlv, Fernandez-Martinez:2023phj}), which we also adopt here. In reality, to fit the observed neutrino oscillation data, we require at least two HNLs in the Type-I Seesaw~\cite{Minkowski:1977sc, Mohapatra:1979ia, Yanagida:1979as,Gell-Mann:1979vob} or its variants like the inverse~\cite{Mohapatra:1986aw,Mohapatra:1986bd}, linear~\cite{Wyler:1982dd, Malinsky:2005bi} or radiative~\cite{Pilaftsis:1991ug,Dev:2012sg} seesaw mechanisms. As long as two HNLs are quasi-degenerate, which is indeed the case with many of the scenarios mentioned above, or if only one HNL is kinematically accessible, our results derived using a single HNL hypothesis remain valid. Similarly, our case (a) with equal flavor mixing can correctly reproduces the observed 3-neutrino flavor mixing data in the inverted-ordering scenario, while our case (b) with $\tau$-flavor dominance is close to the $3\sigma$-allowed range of the normal-ordering scenario which allows up to 90\% $\tau$-component~\cite{Chrzaszcz:2019inj, Drewes:2022akb}. 

From Fig.~\ref{fig:BR_UT_1}, we see that for $m_N<m_\pi$, the dominant decay mode is the invisible $\nu\bar{\nu}\nu$ channel, whereas for $m_\pi<m_N\lesssim 1$ GeV, the 2-body semileptonic pion decay mode $\nu\pi^0$ is the dominant one. For $m_N\gtrsim 1$ GeV, the decays to 3 or more hadrons open up and become significant. 
Decays to $D$ and $B$ mesons are subdominant. For our subsequent analysis, we focus on  case (b) where the HNL mixes only with 
the $\tau$ flavor. 
The leading channels for our study are $N\to \nu\pi^0$ and $N\to \nu e^+e^-$ below $m_{N} < 2$ GeV. These channels are most appropriate for our study because their final states are suitable for reconstruction in fixed-target neutrino experiments. For HNLs with masses $m_{N} \gtrsim 2$ GeV, the HNLs can decay to multi-meson final states. Though more challenging, it is still conceivable to identify the signal. For our work, we focus on the following HNL mass parameter region for the remainder of the presentation: 
\begin{equation}
0.01 \text{ GeV} \leq m_N \leq m_{Z'}/2 \, .
\label{eq:range}
\end{equation}
For the interested reader, we present a more thorough treatment of the relevant HNL partial decay modes in this mass range in Appendix \ref{app: HNL Decays}.

A lot of work has been done in the context of HNL constraints arising from searches in collider experiments, fixed-target experiments, neutrino experiments, nuclear decays, and even cosmological and astrophysical observables; for reviews, see e.g.~Refs.~\cite{Atre:2009rg, Deppisch:2015qwa, deGouvea:2015euy, Cai:2017mow, Bolton:2019pcu, Abdullahi:2022jlv, Fernandez-Martinez:2023phj}. The constraints are usually reported in the HNL mass-mixing plane $(m_N,U_\alpha)$. The  most stringent constraints are in the electron and muon flavors, whereas the tau sector is relatively less constrained, which is the main reason why we focus on the tau sector. The current limits on the HNL mixing versus its mass are taken from Refs.~\cite{bolton, hostert}. For the HNL mass range in~\eqref{eq:range}  relevant to the fixed-target experiments considered here, the relevant limits on $U_{\tau}$ mostly come from CHARM~\cite{Orloff:2002de, Boiarska:2021yho}, BEBC~\cite{WA66:1985mfx, Barouki:2022bkt},  ArgoNeuT~\cite{ArgoNeuT:2021clc} and BaBar~\cite{BaBar:2022cqj}. 
%%%%%%%%%%%%%%%%%%%%%%%%%%%%%%%%%%%%
\section{Benchmark Experiments}
\label{sec:Benchmark}

As we will show below, fixed-target experiments allow us to probe a wide region of HNL parameter space via Drell-Yan production. Numerous neutrino and beam-dump-type fixed-target facilities are currently operating or planned. In this work, we focus on several benchmark experiments with proton beam energies spanning from $\mathcal{O}(10)$ GeV to $\mathcal{O}(400)$ GeV, including SBND~\cite{MicroBooNE:2015bmn}, DarkQuest~\cite{Apyan:2022tsd}, DUNE~\cite{DUNE:2021tad}, and SHiP~\cite{SHiP:2021nfo}. For completeness, we describe below some of the key specifications of these experimental setups.   
\begin{itemize}
    \item[$\bullet$] SBND~\cite{MicroBooNE:2015bmn}: The Short-Baseline Near Detector (SBND) is a currently running LArTPC detector located on the Booster Neutrino Beam (BNB) at Fermilab. An $8~\mathrm{GeV}$ proton beam from the Booster strikes a beryllium target, producing a predominantly $\nu_\mu$ beam from meson decays-in-flight. The detector sits about $L \simeq 110~ \mathrm{m}$ downstream of the target and near the beam axis, with a small transverse offset of $0.74~\mathrm{m}$. The active LArTPC volume is approximately $4~\mathrm{m}$ (width) $\times~4~\mathrm{m}$ (height) $\times~5~\mathrm{m}$ (beam direction), corresponding to an active mass of $\sim 100-110~\mathrm{t}$. For our sensitivity estimates, we take the total protons-on-target (POT) exposure to be $10^{21}$ assuming 3 years of running.

    \item[$\bullet$] DarkQuest~\cite{Apyan:2022tsd}: DarkQuest is a proposed fixed-target proton beam-dump experiment based on the existing SeaQuest/SpinQuest spectrometer on the Fermilab Main Injector (FMI) neutrino-muon line. A high-intensity $120~\mathrm{GeV}$ proton beam impinges on a thin nuclear target located about $1~\mathrm{m}$ upstream of a $5~\mathrm{m}$-long solid-iron dipole magnet (FMAG), which both focuses secondaries and acts as an effective beam dump for non-interacting protons. Downstream of the FMAG, an additional open-aperture dipole (KMAG) and a system of tracking stations and muon identification planes form a compact forward spectrometer. The fiducial decay region for long-lived particles is taken to lie a few meters downstream of FMAG, with decay products required to pass through three $\sim 2~\mathrm{m} \times 2~\mathrm{m}$ tracking stations located at $L \simeq 6, ~13$, and $18.5~\mathrm{m}$ from the target. DarkQuest operates essentially on-axis with respect to the proton beam. For our HNL sensitivity projections, we assume total POT exposures corresponding to the benchmark ``Phase~I'' scenario with a total POT exposure of $10^{18}$ across 2 years of running.

\item[$\bullet$] DUNE~\cite{DUNE:2021tad}: The proposed Deep Underground Neutrino Experiment Near Detector (DUNE ND) complex is placed on the LBNF neutrino beam produced by $120~\mathrm{GeV}$ protons from the Fermilab Main Injector hitting a graphite target. The near hall is located about $L \simeq 574~\mathrm{m}$ downstream of the production target and is roughly $60~\mathrm{m}$ underground. The primary LAr detector (ND-LAr) is a highly segmented LArTPC with pixel-based readout and a fiducial mass of $\sim 50~\mathrm{t}$, with a characteristic active volume of order $4~\mathrm{m} \times 3~\mathrm{m} \times 5~\mathrm{m}$. ND-LAr can be moved transversely in the DUNE-PRISM configuration, sampling off-axis angles at the few-mrad level and effectively scanning different neutrino energy spectra while remaining close to the nominal beam axis. In our numerical analysis, we consider the on-axis configuration, adopting $7\times 10^{21}$ as the total POT exposure for the DUNE ND assuming 7 years of running. 

    \item[$\bullet$] SHiP~\cite{SHiP:2021nfo}: The Search for Hidden Particle (SHiP) is a proposed high-intensity proton beam-dump facility at the CERN Super Proton Synchrotron (SPS). A $400~\mathrm{GeV}$ proton beam is extracted and dumped on a high-density hybrid target, followed by a hadron absorber and an active muon-shield system that strongly suppresses the downstream muon flux. The target consists of an array of molybdenum disks followed by tungsten disks, but the majority of proton interactions occur with the molybdenum. The hidden-sector detector consists of a long evacuated decay volume, approximately $50~\mathrm{m}$ in length with an active transverse acceptance of order $1~\mathrm{m} \times 2.7~\mathrm{m}$, starting about $33~\mathrm{m}$ downstream of the target. The decay volume is followed by a magnetic spectrometer, calorimetry, and a muon system to reconstruct and identify the visible decay products of long-lived states; background taggers surrounding the vessel and a precision timing detector provide additional rejection of neutrino- and muon-induced backgrounds. SHiP is designed to collect a total of $\sim 6\times 10^{20}$ POT over 15 years of operation, with the detector located on-axis with respect to the proton beam.
    
\end{itemize}
 As is evident from the above description, each experimental scenario has a unique combination of beam energy, geometrical setup, and POT. We summarize the key characteristics of these detectors in Table \ref{ExpTables}. The listed dimensions correspond to the sizes of the detector active volumes, while the total POT values in the last column are those assumed throughout our analysis.

 We also considered ICARUS~\cite{ICARUS:2023gpo} which is on-axis to the BNB (further away from SBND) and off-axis to NuMI. Its sensitivity turns out to be weaker than that of SBND, so we do not include it here.

\begin{table}[t!]
\centering
\setlength{\tabcolsep}{2.3pt} 
\renewcommand{\arraystretch}{1.3} 
\footnotesize
\begin{tabularx}{1\linewidth}{l c c c ||c c c c X}
\toprule
Detectors (beam target) & Beam, $E$ [GeV] & Baseline [m] & POT & $l_{\text{Target}}$ [cm]& $\lambda_{\rm int}$ [cm] & $\rho$ [g$\cdot\text{cm}^{-3}$] & $\sigma_{pN}~ [\text{mb}]$ \\
\midrule
SBND (Be)~\cite{MicroBooNE:2015bmn, MiniBooNE:2008hfu} & BNB, 8   & 110 & $1\times10^{21}$ & $71.1$ & $42.10$  & $1.848$ & $21.37$ \\
DarkQuest (Fe)~\cite{Apyan:2022tsd} & FMI, 120 & 5 & $1\times10^{18}$ & $500$ & $16.77$ & $7.87$ & $12.55$  \\
DUNE ND (Graphite)~\cite{DUNE:2021tad, Back:2021yia}   & LBNF, 120 & 574    & $7\times10^{21}$ & $150$& $38.83$ & $2.210$ & $19.36$ \\
SHiP (Mo)~\cite{SHiP:2021nfo} & SPS, 400 & 33 & $6 \times 10^{20}$ & $140$& $15.25$ & $10.2$ & $10.68$ \\
\bottomrule
\end{tabularx}
\caption{Key detector specifications, beam-target elemental data, and inelastic hadronic cross sections for our benchmark experiments. The POT presented is the total POT assumed in this study.}
\label{ExpTables}
\end{table}

%%%%%%%%%%%%%%%%%%%%%%%%%%%%%%%%%%%
\section{Signals and Backgrounds}
\label{sec:SB} 

\subsection{Signal Event Calculation}
\label{sec:signal}

In fixed-target or beam-dump experiments, one relies on the large number of incoming protons to hit the target, which is often measured by POT. The predicted BSM signal rate under consideration depends on the production cross section of a mediator particle 
$X$, $\sigma(pp\to X)$  and its decay branching fraction to the observable SM final state ${\rm BR}(X\to {\rm final\ state})$ with detection efficiency $\text{eff}_f$. 
The number of signal events for fixed-target experiments are given by~\cite{Batell:2020vqn}
\begin{equation}
N_{\text{Detected}} = \text{POT}\times \frac{\sigma(pp\to X)}{\sigma_{pp}}\times {\rm BR}(X\to {\rm final\  state})\times\text{eff}_f,
\label{eq:events} 
\end{equation}
where $\sigma_{pp}$ represents the total cross sections for all processes allowed in the $pp$ collisions.

In our consideration, the HNL signal comes from the $Z'$ resonant production and its subsequent decay. A light $Z'$ with a mass below $\sim1$ GeV will be mostly produced via bremsstrahlung and/or exotic decays of neutral/charged mesons~\cite{deNiverville:2016rqh,Dutta:2021cip,Dutta:2025fgz}. However, for heavier $Z'$, the meson-production channels become kinematically inaccessible, while the bremsstrahlung process is strongly suppressed by phase space. Therefore, the resonant production via the $q \bar q$ annihilation, namely the ``Drell-Yan'' process, stands out and can contribute significantly in the DIS regime.

%%%%%%%%%%%%%%%%%%%%%%%%%%%%%%%%

The incoming $q$ and $\bar{q}$ carry some momentum fractions $x_1$ and $x_2$ of their parent protons. The squared-energy in the partonic center-of-mass frame is $\hat{s} = \tau s = x_1x_2s$. 
The cross section of $X$ production in terms of the partonic luminosity  $\mathcal{L}_{i,j}$ can be written as 
\begin{align}
\sigma(pp\to X) &= \int_{\tau_0}^1d\tau\sum_{q,\bar{q}}\frac{d\mathcal{L}_{q,\bar{q}}}{d\tau}\,\hat{\sigma}(q\bar{q}\to X), \quad \tau_0= {m^2_X\over s}, \\
{\rm where}\quad \frac{d\mathcal{L}_{i,j}}{d\tau} &= \frac{1}{1 + \delta_{ij}}\int_{\tau}^1\frac{d\xi}{\xi}\left[f_{i}\big(\xi,Q^2\big)f_j\left(\frac{\tau}{\xi},Q^2\right) + (i\leftrightarrow j)\right] .
\end{align}
where $Q^2$ is the QCD factorization scale, 
$f_{i,j}$ are the PDFs. Generically, the cross section for the $Z'$ resonant  production, including its decay to a fermion pair is
\begin{equation}
    \hat{\sigma}(q\bar{q} \to Z' \to f\bar{f}) = \frac{g_{X}^4}{108\pi}\frac{\beta_f \hat{s}\ \left(g_{V,q}^2 + g_{A,q}^2\right)}{(\hat{s} - m_{Z'}^2)^2 + \Gamma_{Z'}^2 m_{Z'}^2}\left[\left(g_{V,f}^2 + g_{A,f}^2\right)\beta_f^2 + g_{V,f}^2 \frac{6m_f^2}{\hat{s}}\right] , 
\end{equation} 
where $\beta_f = (1 - 4m_f^2/\hat{s})^{1/2}$ and $\Gamma_{Z'}$ is the total decay width of $Z'$, obtained by summing over the partial decay rates given in Eq.~\eqref{eq:width}. The coefficients $g_{V,f}, g_{A,f}$ are model-specific (see Table \ref{ModelCouplings}). %
In the narrow width approximation $\Gamma_{Z'}/m_{Z'} \ll 1$, the propagator simplifies to 
\begin{equation}
\frac{1}{(\hat{s} - m_{Z'}^2)^2 + \Gamma_{Z'}^2m_{Z'}^2} \to \frac{\pi}{\Gamma_{Z'}\ m_{Z'}}\ \delta(\hat{s} - m_{Z'}^2).
\end{equation}
The HNL sub-process cross section in the $U(1)_{B-L}$ model is given by
\begin{equation}
\hat{\sigma}(q\bar{q}\to Z' \to NN) = \frac{g_{B-L}^4}{1944}\frac{\beta_N^3 \tau}{\Gamma_{Z'}m_{Z'}} \delta\left(\tau - {m^2_Z \over s}\right).
\label{eq:NN}
\end{equation}
At the resonance, the sub-process cross section scales as  $g^2_{B-L}/m_{Z'}^2$.
Finally, we arrive at the cross section for $NN$ production via the Drell-Yan process in the $pp$ collisions from the mediator $Z'$ production and decay:
\begin{equation}
\sigma(p p \to Z' \to NN) = \frac{g_{B-L}^4}{1944\Gamma_{Z'}}\frac{\beta^3_N m_{Z'}}{s}\cdot\frac{1}{2}\sum_{q,\bar{q}}\int_{m_{Z'}^2/s}^1\frac{dx_2}{x_2}\left[f_{q}\left(\frac{m_{Z'}^2}{x_2s}, m_{Z'}^2\right)f_{\bar{q}}(x_2, m_{Z'}^2) + (q\leftrightarrow \bar{q})\right] \label{eq:NNxs}
\end{equation}

\begin{figure}[tb]
    \centering  \includegraphics[width=0.6\linewidth]{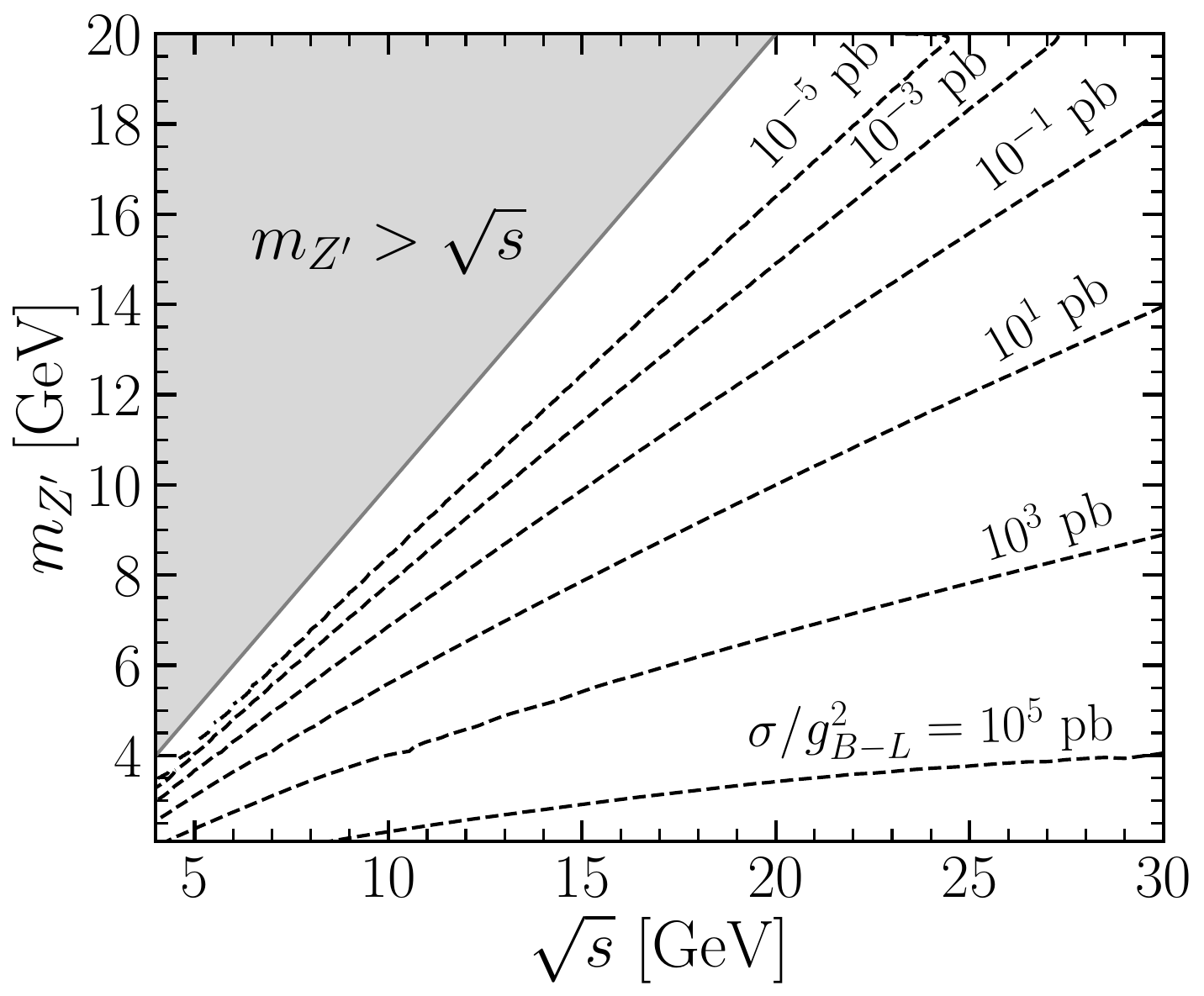}
    \caption{Iso-cross section contours for the process $pp \to Z'\to NN$ normalized to unit coupling $g_{B-L}$ with $m_N \ll m_{Z'}$.
    The minimum scale is fixed at $\sqrt{\hat{s}}_{\rm min} = 2$ GeV. In the gray-shaded region, $m_{Z'}>\sqrt s$,  on-shell production of $Z'$ is kinematically forbidden.
    }
    \label{fig:mZp_Max_Limits}
\end{figure}

We proceed by exploring the regions that lie within the on-shell mass resonance of the $Z'$. We modified the \texttt{B-L-4UFO} \cite{Amrith:2018yfb, Deppisch:2018eth, Basso:2008iv} model file in \texttt{FeynRules}~\cite{Alloul:2013bka} to allow arbitrary couplings to an HNL via $U_e, U_\mu, U_\tau$.  For each experimental setup mentioned earlier, we consider HNL production in $pp$ collisions in the center-of-mass frame. We used the \texttt{CT18qed}~\cite{Xie:2021equ} PDF set for the proton and imposed a minimum partonic center-of-mass energy of $\sqrt{\hat{s}}_{\min}=2~\mathrm{GeV}$.
To assess the validity of the PDF-based perturbative QCD treatment near threshold, we varied the input $\sqrt{\hat{s}}_{\min}$ over the range $1-2~\mathrm{GeV}$ and found that our resonant-production results were not appreciably affected. We then computed the total production cross section for $pp \to Z' \to NN$ as a function of the hadronic center-of-mass energy $\sqrt{s}$. In Fig.~\ref{fig:mZp_Max_Limits}, we show the iso-cross section contours for $\sigma/g^2_{B-L}$ in the $m_{Z'}$-$\sqrt s$ plane. The cross section increases with the colliding energy logarithmically due to the increasing parton luminosity, but falls as an inverse power-law in $Z'$ mass. To accommodate the other $U(1)_X$ benchmark models discussed in Section~\ref{Sec:Model}, one needs to multiply by the appropriate $Z'$ coupling $g_{V,q}^2g_X^2$  for the initial state quarks and rescale by the branching fraction BR$(Z'\to NN)$ for the new model. For example, to compute the cross section for the $U(1)_{B- 3L_{\tau}}$ model with coupling $g_{B-3L_{\tau}}$, charges $g_{V,q}$ and $g_{A,N}$, and total $Z'$ decay width $\Gamma_{B-3L_{\tau}}(g_{B-3L_{\tau}}^2)$, we rescale by taking 
\begin{equation} 
\sigma_{B-3L_\tau} = \sigma_{B-L}\times\frac{g_{B-3L_{\tau}}^4}{g_{B-L}^4}\times\frac{\Gamma_{B-L}(g_{B-L}^2)}{\Gamma_{B-3L_{\tau}}(g_{B-3L_{\tau}}^2)}\times 9g_{V,q}^2g_{A,N}^2.
\end{equation}

In general, a large POT exposure is expected to yield high event rates, offering strong potential for new-physics searches. For a more accurate estimate of the signal yields, we account for proton-beam leakage (or losses) in each facility, which reduces the effective event counts. The leakage percentage for each experiment can be estimated by taking $p_{\text{leak}} \approx   \text{exp}(-l_{\text{target}}/\lambda_{\text{int}})$, where the relevant target length $l_{\text{target}}$ and interaction length $\lambda_{\rm int}$ are taken from Table \ref{ExpTables}. One could use the target geometry and beam energy in a \texttt{GEANT4} simulation~\cite{GEANT4:2002zbu} to more accurately estimate the percentage of protons that leak out of the target~\cite{Dutta:2025ddd}. We find leakages of $\sim18\%$ for SBND, $\sim 4\%$ for DUNE ND, and negligible leakages for DarkQuest and SHiP due to their long target lengths relative to their interaction lengths.

\begin{figure}[tb]
    \centering
\includegraphics[width=0.96\textwidth]{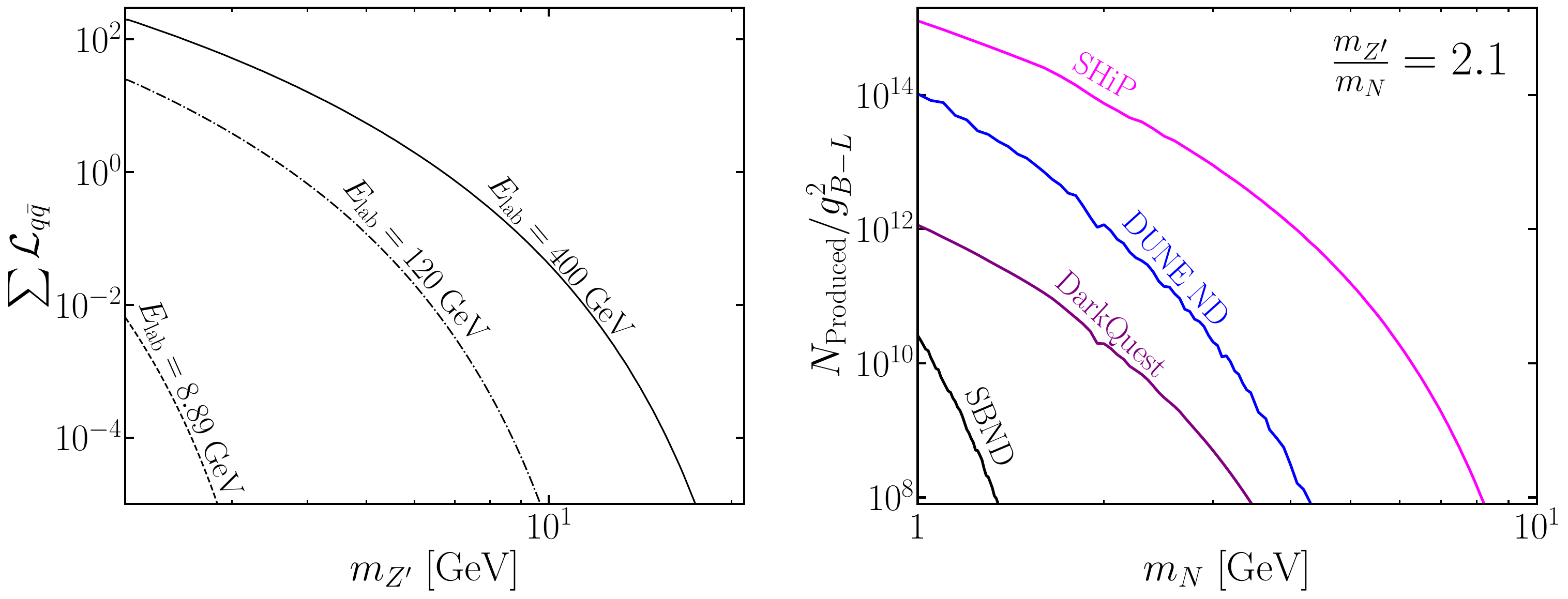}
    \caption{(Left) Quark parton luminosity for on-shell light vector boson production at various laboratory-frame energies for this study.  (Right) The number of HNLs produced within the decay volume in each benchmark experiment, where we have fixed $m_{Z'}/{m_N} = 2.1$.}
    \label{fig:NFlux}
\end{figure}

The total number of HNLs produced through our process for each experiment are:
\begin{equation}
\label{N_PFlux}
N_{\text{Produced}} = \text{POT} \times \Big(1 - p_{\text{leak}}\Big)\times  \frac{\sigma_{\rm BSM}}{\sigma_{pN}}
\end{equation}
We approximate the inelastic hadronic cross section per target nucleon as $\sigma_{pN} \approx (\lambda_{\rm int}\ \rho N_{A})^{-1}$, where $\rho$ is the target mass density and $N_{A}$ is the Avogadro number. $\sigma_{pN}$ for each experiment is provided in Table
\ref{ExpTables}. 
We show the number of produced events, $N_{\text{Produced}}$, as a function of $m_N$ for the various experiments in the right panel of Fig.~\ref{fig:NFlux}. We only present the number of HNLs produced, which are oriented in the direction of the decay volume. For a given POT, this requirement reduces the number of HNLs produced relative to Eq.~\eqref{N_PFlux}, particularly for experiments with longer baselines. 
The corresponding partonic luminosities---shown in the left panel of Fig.~\ref{fig:NFlux}---provide the key ingredient governing the Drell-Yan production rate.

An important consideration for fixed-target experiments is the total detection efficiency of the detector.  In general, for $N_{\text{MC}}$ random Monte-Carlo-generated events, the detection efficiency is calculated by~\cite{Berlin:2018pwi}
\begin{equation}
\text{eff} = m\Gamma\int_{z_{\text{min}}}^{z_{\text{max}}}dz\sum_{\text{events}~\in~ \text{geom.}}\frac{e^{-z(m/p_z)\Gamma}}{N_{\text{MC}}\ p_z},
\label{eq:eff}
\end{equation}
where $z_{\min}$ and $z_{\max}$ define the length scale of the ``effective'' decay volume along the beam axis, with the corresponding component of the decaying particle's momentum $p_z$.
For cases such as the SBND and DUNE ND where the HNL decays within the detector, the sum of the events within the geometry is independent of the decay depth, so we are free to take the integral and find: 
\begin{equation}
\text{eff} = \frac{1}{N_{\rm MC}}\sum_{\text{events}~\in~\text{geom.}}\big(e^{-z_{\text{min}}(m/p_z)\Gamma} - e^{-z_{\text{max}}(m/p_z)\Gamma}\big).
\end{equation}
In this case, we model the geometry by an ``acceptance cone'', whose base is set by the detector length and width. The cone is characterized by acceptance angles $\theta_w$ along the $x$-axis and $\theta_h$ along the $y$-axis defined by the decay volume and the fiducial volume of the detector. For DarkQuest and SHiP, the HNL can decay within the decay volume  and then the decay products must propagate to the ECAL detector on the other end of the decay chamber, so the integration/summation order in Eq.~\eqref{eq:eff} cannot be swapped and the efficiency is calculated numerically.

Taking all of these considerations into account, the number of events which reach the detector are given by \cite{Batell:2016zod}
\begin{equation}
    N_{\text{Detected}} = N_{\text{Produced}}\times {\rm BR}(N\to f) \times~\text{eff}_f , 
\end{equation}
where
$\text{eff}_f$  is the efficiency factor for $N$ to  decay to an observable final state $f$ with a branching fraction ${\rm BR}(N\to f)$.
%
%%%%%%%%%%%%%%%%%%%%%%%%%%%%%%%%%%%%%%%
\subsection{Background Considerations}
\label{sec:backgrounds}
The resonant production of our HNLs results in a kinematic separation between our signal and background. 
One of the important features for our signal is the highly-boosted kinematics along the high-energy beam direction. We thus expect to have energetic decay products with a  smaller opening angle with respect to the beam direction to distinguish them from the background processes. 
In our analysis, we focus on the following clean decay channels: 
\begin{equation}
    N\to \ \nu_{\tau}e^+e^-\quad {\rm and}\quad \nu_{\tau}\pi^0.
    \label{eq:signal}
\end{equation}
We take advantage of the unique signal  kinematics and search for energetic $e^\pm$ and $\pi^0\to \gamma\gamma$. 
In our estimation of the signal sensitivity, we adopt a measure of 
the $90\%$ Confidence Level (C.L.) exclusion limits. 
For the case of nonzero background events, we estimate the sensitivity by computing the significance with the profile-likelihood approximation described in~\cite{ParticleDataGroup:2024cfk}
\begin{equation}
Z_0 = \sqrt{2\left[(S + B)\ln\left({1 + \frac{S}{B}}\right) - S\right]},
\label{eq:z0} 
\end{equation}  
where $S$ and $B$ are respectively the number of events for our signal and the background. The $90\%$ C.L.~sensitivity corresponds to $Z_0 = 1.64$~\cite{ParticleDataGroup:2024cfk}. 
We include a $10\%$ systematic uncertainty on the background estimation $B\to 1.1 B$. 
For optimistic cases of the background-free channels, $90\%$ C.L.~corresponds to $S= 2.3$ signal events~\cite{ParticleDataGroup:2024cfk}. 
\begin{figure}[tb]
    \centering
\includegraphics[width=0.5\linewidth]{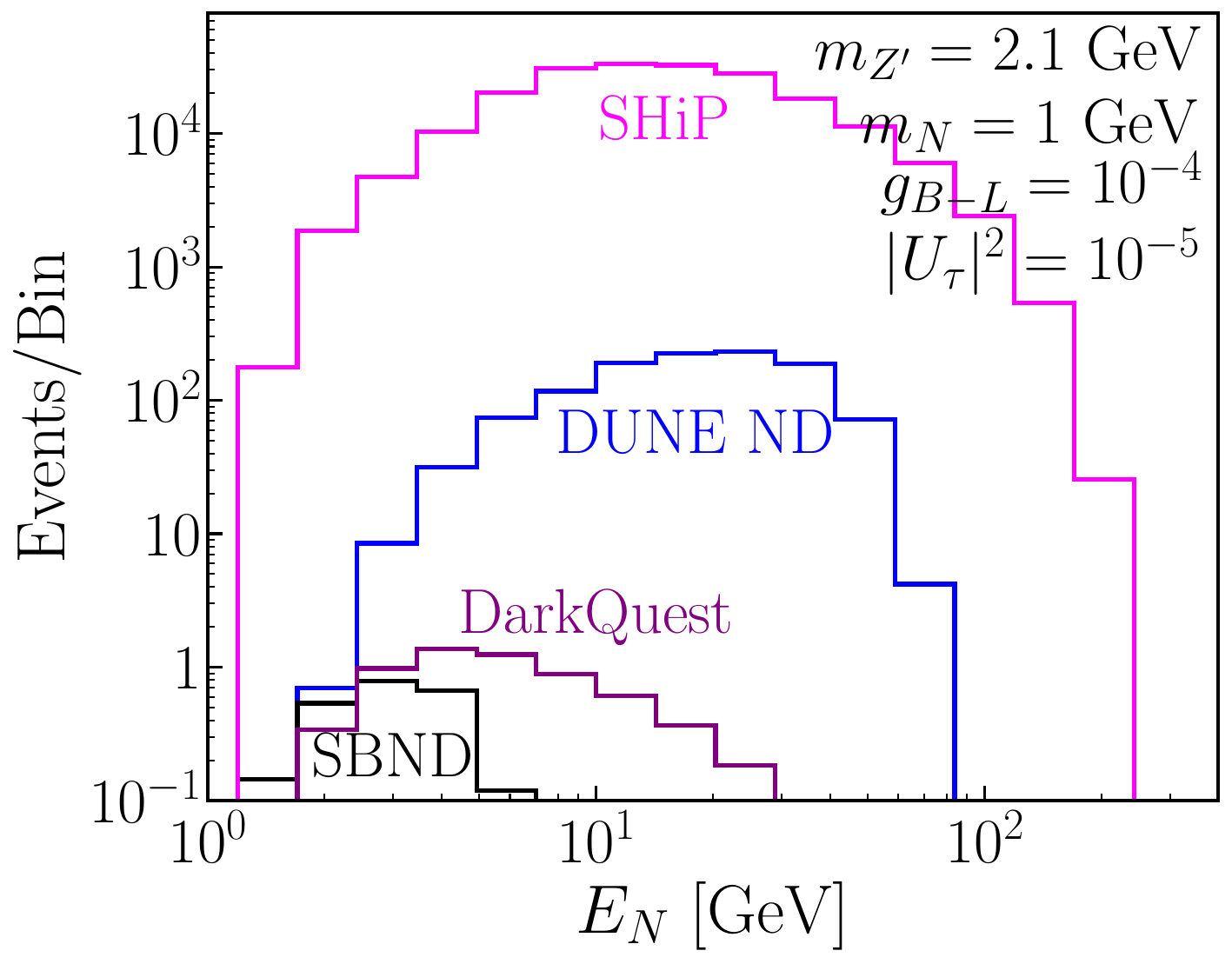} \\
   \caption{Kinematic distributions for $N$
    energy at various experiments. 
}
   \label{fig:mN_Event_Distributions}
\end{figure} 
%
%

%%%%%%%%%%%%%%%%%%%%%%%%%%%%%
\subsubsection{SBND}

The large flux of $\nu_\mu$ provides  the primary source for physics studies at SBND. The neutrino-induced background, on the other hand, is the primary concern for HNL searches. 
For the signals as in Eq.~(\ref{eq:signal}), the neutral-current $\nu$-nucleus interactions producing $\pi^0\to\gamma\gamma$ yield the main backgrounds, where
the diphoton pair can be misidentified as an $e^+e^-$ pair. On the other hand, the backgrounds lead to soft photons. In contrast, Fig.~\ref{fig:mN_Event_Distributions} shows the energy distribution of the signal $N$ at SBND and other experiments. Depending on the available beam energies as well as the mass scales $m_{Z'}$ and $m_N$, the signal energy spectrum is significantly harder than the SM background processes. Indeed, recent results~\cite{sbndpi0, sbndee} suggest that the backgrounds for each channel are negligible provided that we take the following minimum energy cuts 
\begin{equation}
E_{e^+e^-}, \ 
E_{\pi^0} > 1.8~\text{GeV}.
\end{equation}
We thus consider SBND to provide a background-free experiment for the signals under our consideration. 

\subsubsection{DarkQuest}

For DarkQuest, neutrino-induced backgrounds are suppressed since most of the charged mesons sourcing neutrinos are absorbed. Once again, we exploit the energetic nature of the Drell-Yan production process to reduce these backgrounds further. The energy distribution of the HNLs is shown in Fig.~\ref{fig:mN_Event_Distributions}. We find that the total background energies from proton scattering in the target and FMAG, which can successfully pass through stations 1, 2, and 3 to reach the ECAL, are very small. Therefore, if we take minimum energy cuts~\cite{Apyan:2022tsd,Blinov:2024gcw,Dutta:2024nhg}
\begin{equation}
E_{e^+e^-} > 500\text{ MeV}\quad \text{and}\quad E_{\pi^0} > 2 \text{ GeV},
\end{equation}
then the background contribution becomes negligible. 

\subsubsection{DUNE ND}

\begin{figure}[tb]
    \centering
    \includegraphics[width=0.495\linewidth]{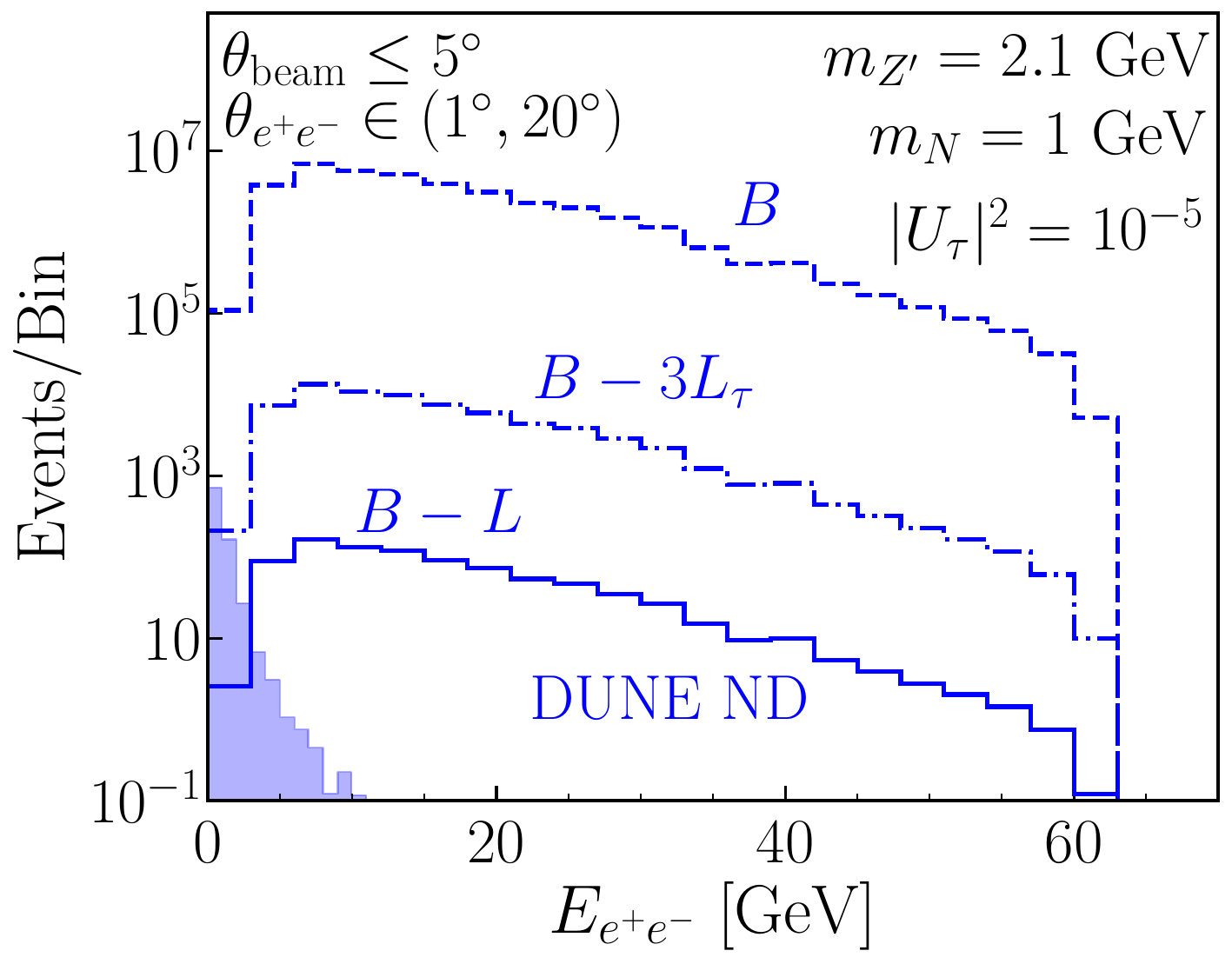} 
     \includegraphics[width=0.48\linewidth]{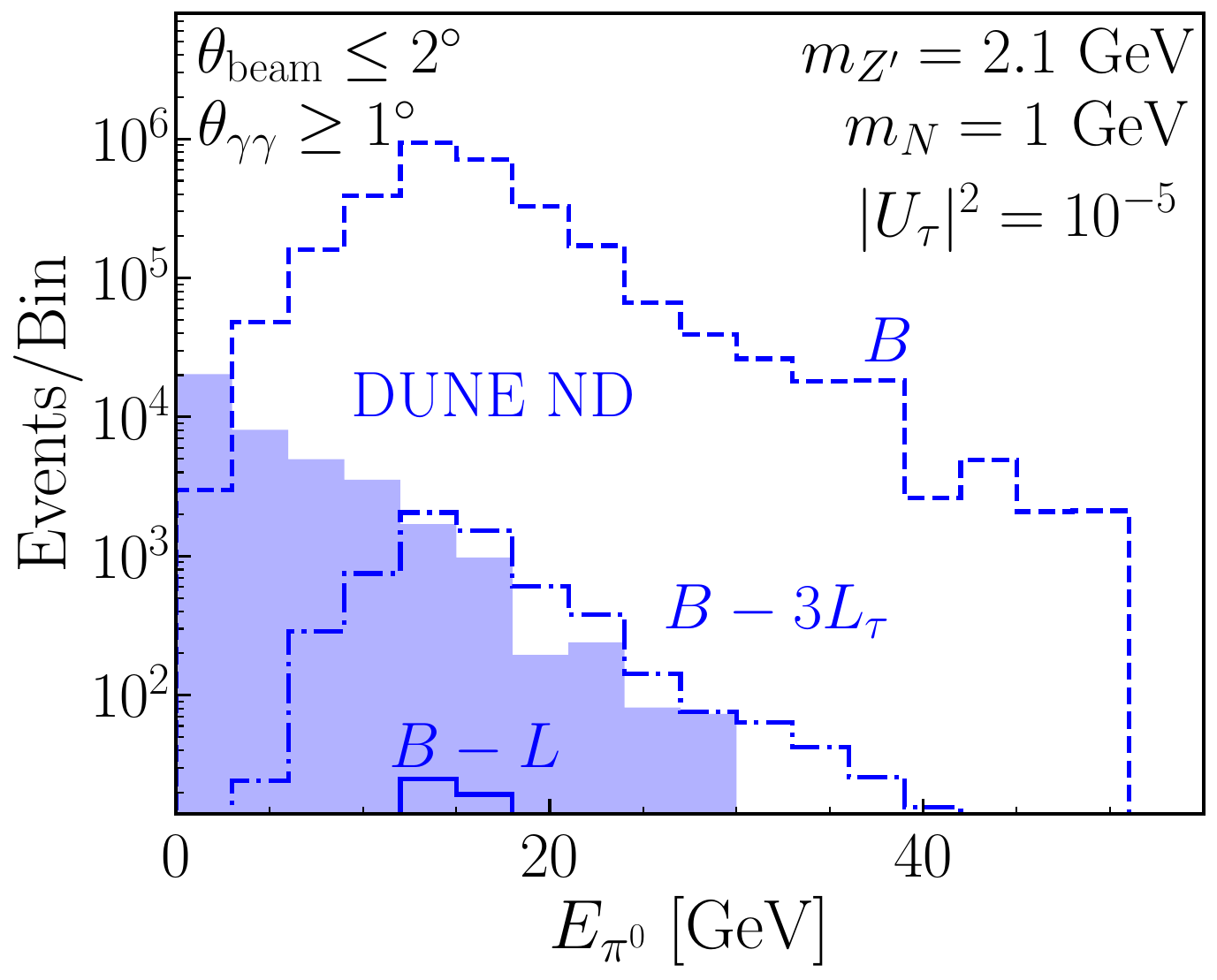}
   \caption{Kinematic distributions for $N$ decay products 
(a) energy of $e^+e^-$ pair, (b) and the $\pi^0$ energy at DUNE ND, for the $U(1)_{B-L}$ (solid curve), $U(1)_{B-3L_{\tau}}$ (dashed curve) and $U(1)_B$ (dotted curve) models with benchmark couplings as in Eq.~(\ref{eq:coup}). 
}
   \label{fig:mN_Event_Distributions_DUNE}
\end{figure}
At the DUNE ND, the $e^+e^-$ (with zero proton) background from neutrino scattering has been studied in detail in Ref.~\cite{Brdar:2025hqi}. It arises from $\pi^0$ decay via $\pi^0\rightarrow\gamma e^+e^-$, where the final-state photon has energy below the reconstruction threshold, as well as from events in which $\gamma e^\pm$ final states are misidentified as  $e^+e^-$ pairs due to photon-electron misidentification (at the $\sim$ 18\%  level). 
Additionally, the $e^+e^-$ pair arising from HNLs should be boosted forward along the beamline. Therefore, we impose the following acceptance cuts~\cite{Dutta:2024nhg,Brdar:2025hqi} on the angle $\theta_{\text{beam}}$ with respect to the beam axis and on the opening angle $\theta_{ee}$ of the $e^+e^-$ pair: 
\begin{equation}
E_{e^+ e^-} > 5~\rm{GeV}, \quad 
\theta_{\text{beam}} \leq 5^{\circ}\quad\text{and}\quad 1^{\circ} < \theta_{ee} < 20^{\circ}.
\end{equation}
We show the $e^+e^-$ pair energy distributions for the SM background expectation (shaded region) and the signal for the three representative models in Fig.~\ref{fig:mN_Event_Distributions_DUNE}(a). The magnitude difference among the three models is primarily due to the benchmark choice of the couplings motivated by the current existing bounds as in Eq.~(\ref{eq:coup}), although there are also minor model-dependent factors. We observe a significant difference between the signal and the background. This result is due to heavy particle decays of the $Z'$ and $N$ signal, along with the boost in high-energy collisions. We find that the background can be essentially eliminated, so we take our signal in the $N\to \nu e^+e^-$ channel to be background-free.

For the other signal channel $N\to \nu\pi^0$,  there is a sizable albeit isotropic pion background arising from neutral-current neutrino-nucleus scattering. However, the kinematic features discussed above are also present. Figure \ref{fig:mN_Event_Distributions_DUNE}(b) illustrates signal and background distributions of the $\pi^0$, with the same benchmark choice of the couplings as in Eq.~(\ref{eq:coup}). 
We target the pions produced near the beamline in our signal by imposing the following optimal cuts on the neutral pion energy $E_{\pi^0}$, the angle $\theta_{\text{beam}}$ with respect to the beam axis, and the opening angle $\theta_{\gamma\gamma}$ between the photon-photon pair:
\begin{equation}
E_{\pi^0} > 20~\text{GeV}, \quad \theta_{\text{beam}} < 2^{\circ},\quad \text{and} \quad \theta_{\gamma\gamma} \geq 1^{\circ} .
\end{equation}
Given the potential uncertainty of the background estimation \cite{Brdar:2025hqi}, we adopt a $10\%$ systematic uncertainty on the background estimate by taking $B_{\rm est.} = 1.1 B$. With these acceptance cuts, we are able to optimize the signal-to-background ratio and evaluate the signal sensitivity using Eq.~(\ref{eq:z0}). 

\subsubsection{SHiP}

For the SHiP experiment, the main source of backgrounds is muon and neutrino-induced interactions within the detector or the surrounding material.  However, due to the decay vessel maintaining a pressure of $< 10^{-2}$ bar to  suppress neutrino-air interactions, the background environments compared to our signal are negligible ~\cite{SHiP:2021nfo}. Therefore, in our analysis, we will not require any stringent acceptance cuts for SHiP.

\begin{figure}[tb]
   \centering
    \includegraphics[scale = 0.21]{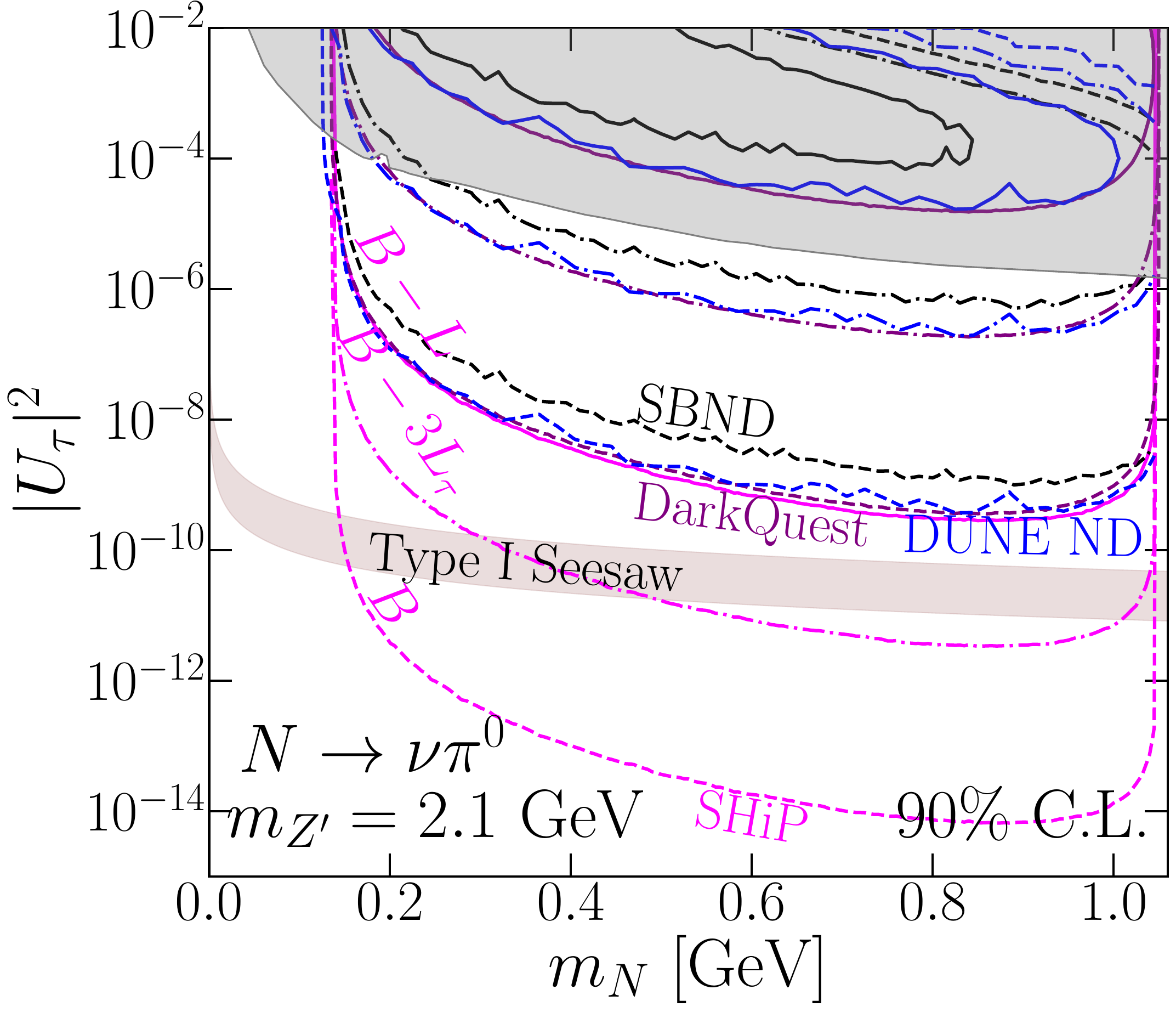}
    \includegraphics[scale = 0.21]{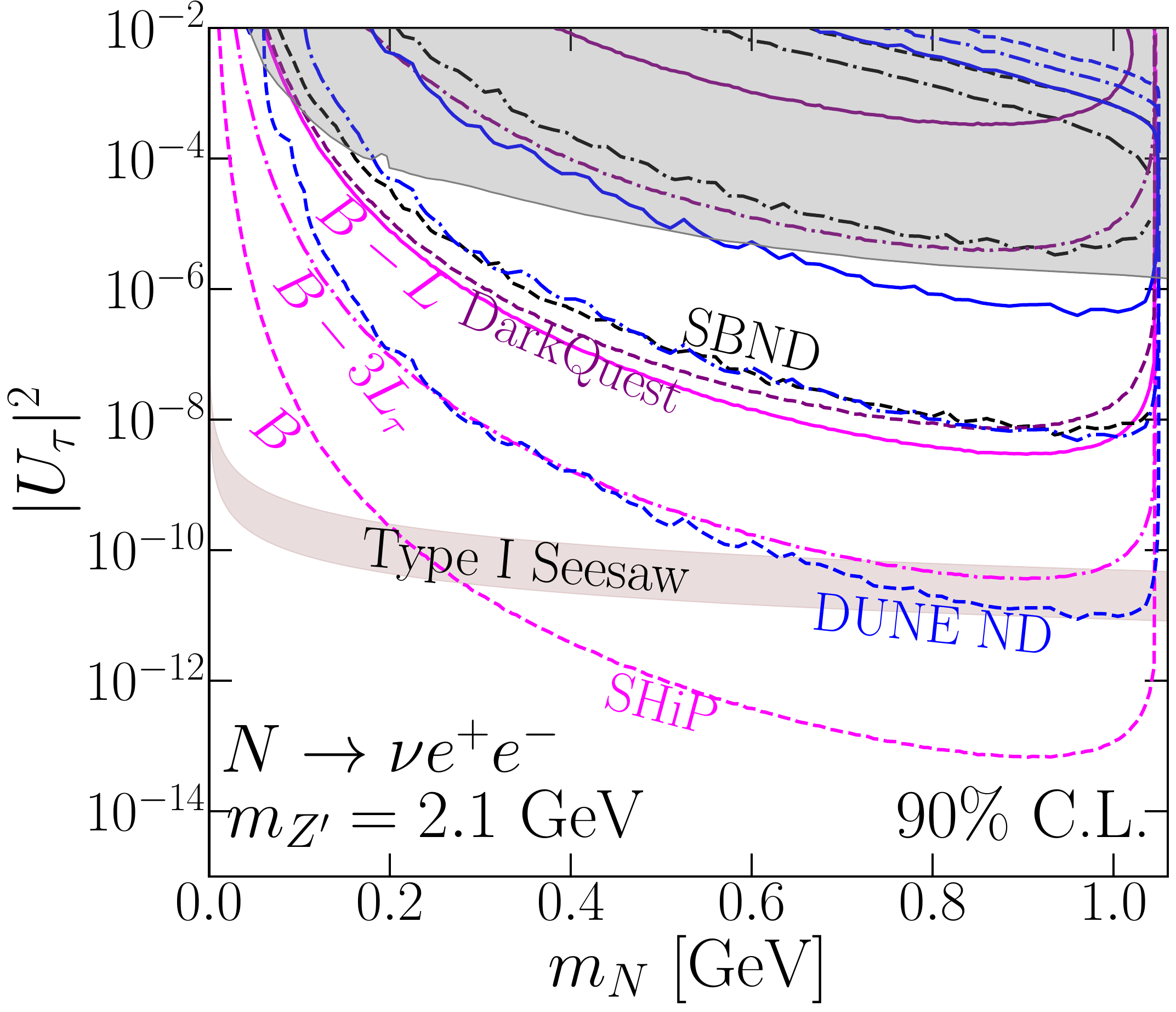}
    \includegraphics[scale = 0.21]{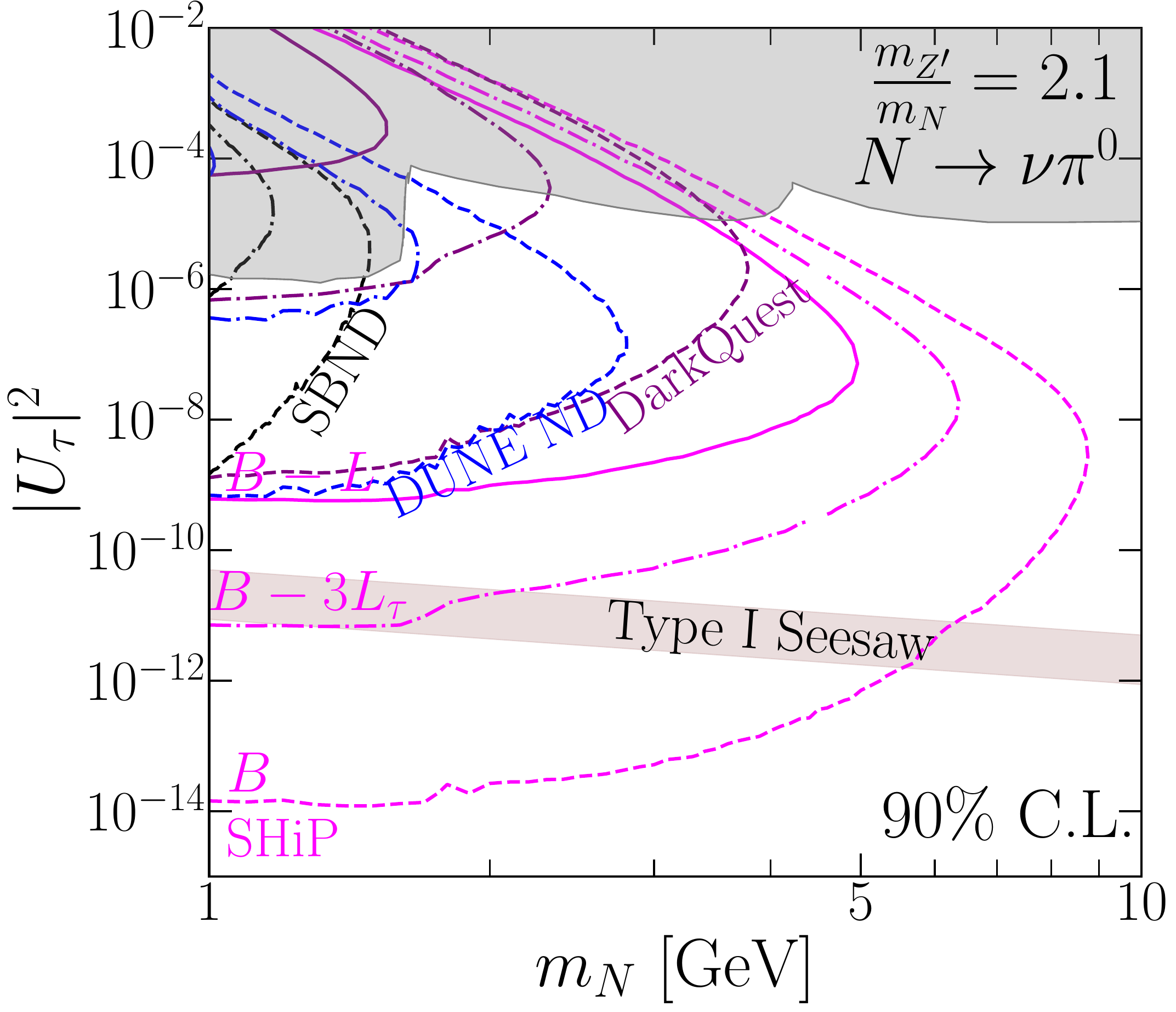}
    \includegraphics[scale = 0.21]{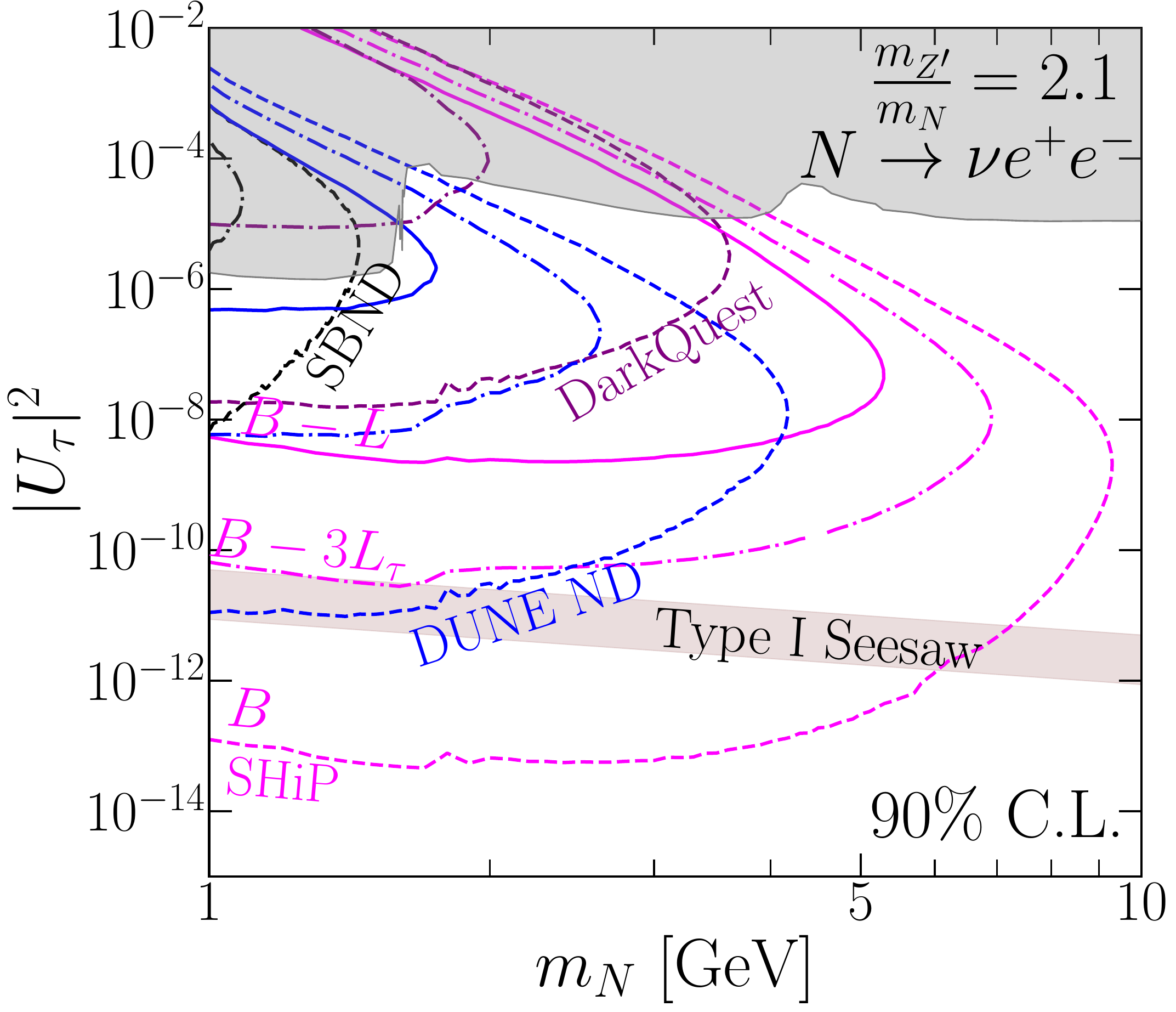}
    \caption{Sensitivity contours in $|U_\tau|^2-m_N$ plane at $90\%$ C.L. for fixed $m_{Z'} = 2.1$ GeV in the (a) $N\to \nu\pi^0$ and (b) $N\to \nu e^+e^-$ channels and with fixed ratio $m_{Z'}/m_N = 2.1$ for the (c) $N\to \nu \pi^0$ and (d) $N\to \nu e^+e^-$ channels, for the SBND (black), DarkQuest (purple), DUNE ND (blue), and SHiP (magenta) experiments. The solid, dotted, and dashed curves are for the $U(1)_{B-L}$, $U(1)_{B-3L_{\tau}}$, and $U(1)_B$ models respectively for the appropriate benchmark couplings in Eq.~(\ref{eq:coup}). The upper gray-shaded region represents the model-independent neutrino mass-mixing limits from Refs.~\cite{bolton, hostert}. The lightly shaded band shows the canonical Type-I seesaw prediction to get the correct atmospheric (upper line) and solar (lower line) mass-squared splittings.}
    \centering
    \label{fig:Sensitivity_pinu}
\end{figure}

\begin{figure}[tb]
   \centering
    \includegraphics[scale = 0.27]{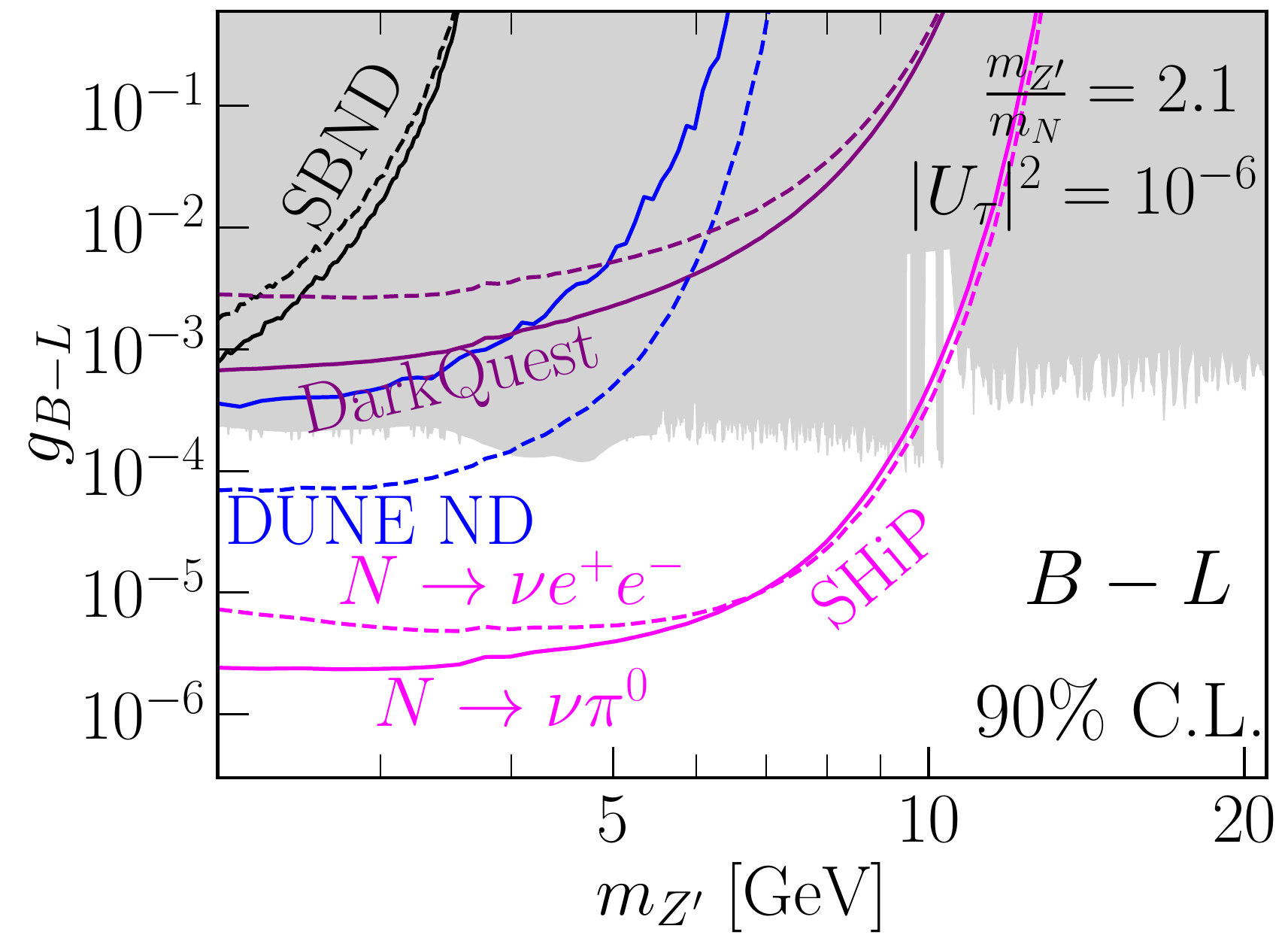}
    \includegraphics[scale = 0.27]{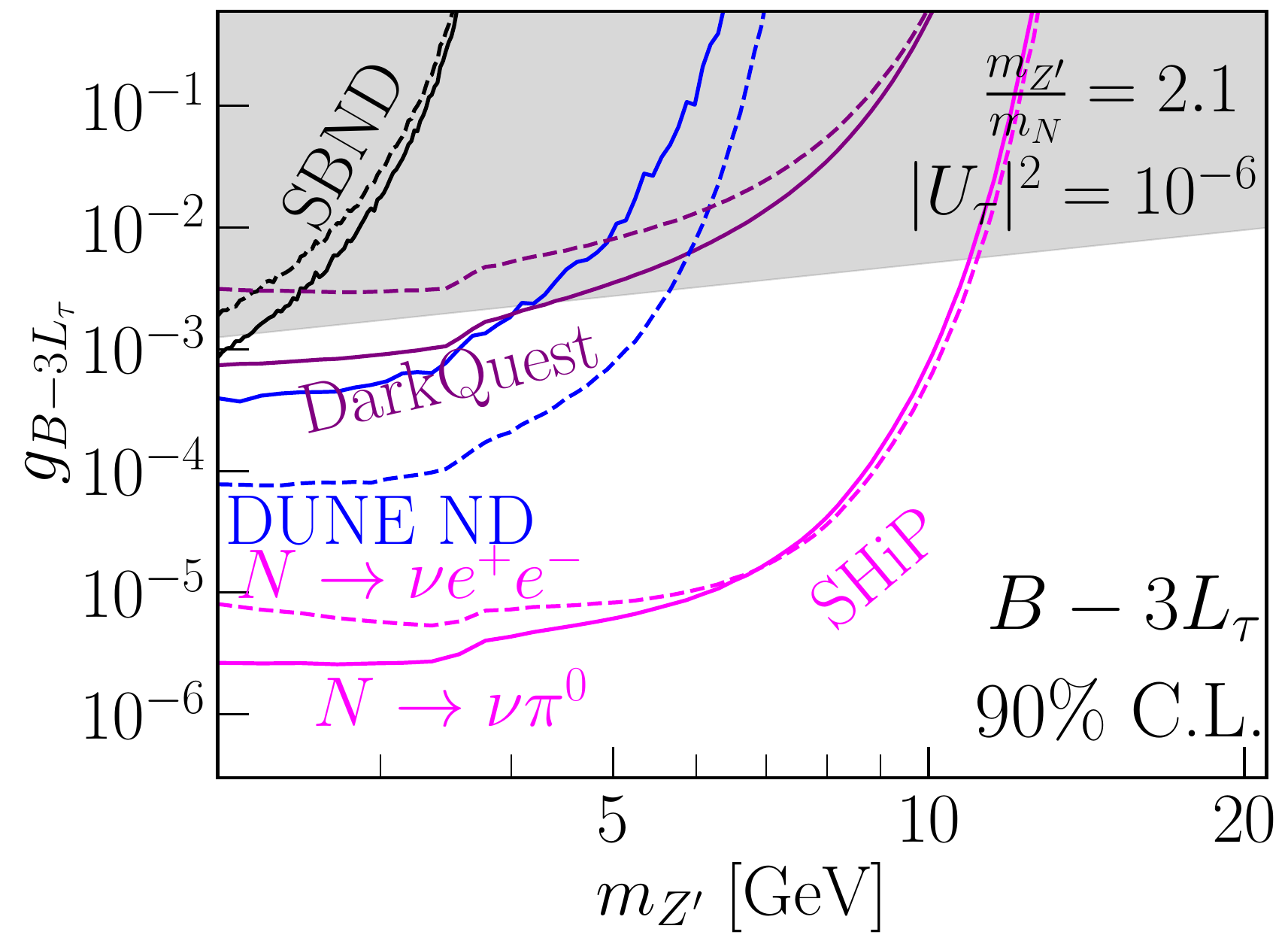}
    \includegraphics[scale = 0.27]{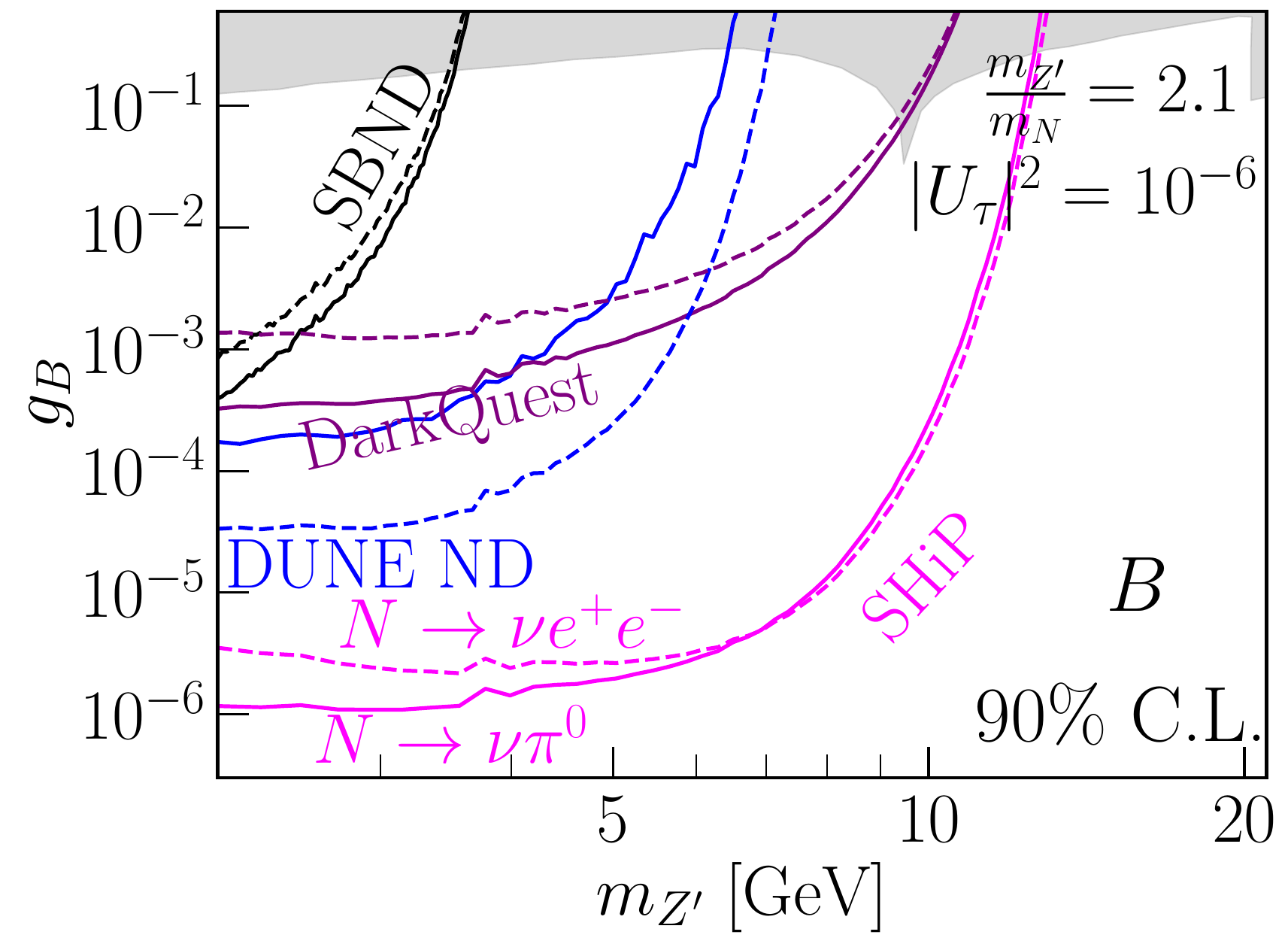}
    \caption{Sensitivity contours in $g_X-m_{Z'}$ plane at $90\%$ C.L.~with $N\to \nu \pi^0$ (solid) and $N\to \nu e^+ e^-$ (dashed) for the three $Z'$ models (a) $U(1)_{B-L},\ \text{(b) } U(1)_{B-3\tau}$ and (c) $U(1)_B$, at the four experiments SBND (black), DUNE ND (blue), DarkQuest (purple), and SHiP (magenta). The gray-shaded regions represent the model-dependent bounds described earlier in Sec.~\ref{Sec:Model}, taken from Refs.~\cite{Bauer:2018onh, Ilten:2018crw}.
    }
    \centering
    \label{fig:SensitivityZ}
\end{figure}

\section{Sensitivity Reach}
\label{Sensitivity}

After applying the optimal cuts, we present the resulting HNL signal sensitivities at the 90\% C.L. in the $|U_\tau|^2$–$m_N$ plane for the four experiments and the three benchmark models in Fig.~\ref{fig:Sensitivity_pinu}. Again, the magnitude difference among the models is primarily due to the benchmark choice of the couplings as in Eq.~(\ref{eq:coup}), although there are also minor model-dependent factors.
The upper panels are for fixed mass $m_{Z'}=2.1$ GeV, whereas the lower panels present the cases with a fixed mass ratio  $m_{Z'}/m_N=2.1$. The left (right) panels are for the  $\nu\pi^0$ ($\nu e^+e^-$) channels with the same benchmark coupling choices as in Eq.~(\ref{eq:coup}). The upper gray-shaded region represents the model-independent neutrino mass-mixing limits coming from CHARM~\cite{Boiarska:2021yho}, BEBC~\cite{Barouki:2022bkt} and ArgoNeuT~\cite{ArgoNeuT:2021clc}. The lower light-brown band is the target region predicted by the canonical Type-I Seesaw mass mixing relation $|U|^2\sim m_{\nu}/m_N$~\cite{Minkowski:1977sc, Mohapatra:1979ia, Yanagida:1979as, Gell-Mann:1979vob}, where we have taken $\sqrt{\Delta m^2_{\rm atm}}\simeq 50$ meV (atmospheric) and $\sqrt{\Delta m^2_{\rm sol}}\simeq 8.7$ meV (solar)~\cite{Esteban:2024eli} 
as the representative values for $m_\nu$ to get the upper and lower edge of the band, respectively. We find that the $\nu \pi^0$ channel is more sensitive than the $\nu e^+e^-$ channel for SBND, DarkQuest, and SHiP. For DUNE ND, the background presence in the $\nu\pi^0$ channel results in the $\nu e^+ e^-$ channel having a better sensitivity to our process. 
In Figs.~\ref{fig:Sensitivity_pinu}(a) and \ref{fig:Sensitivity_pinu}(c) for the $\nu\pi^0$ channel, we see that SBND and DarkQuest can improve the reach to as low as $|U_{\tau}| \approx 3\times 10^{-4}$ for the $U(1)_B$ benchmark. From Figs.~\ref{fig:Sensitivity_pinu}(b) and \ref{fig:Sensitivity_pinu}(d) for the $\nu e^+ e^-$ channel, DUNE ND can improve the reach to $|U_\tau|\approx 3\times 10^{-6}$ for the $U(1)_B$ benchmark. SHiP can achieve substantial sensitivity in the same figures, reaching
$|U_\tau|\approx 10^{-7}$ in the kinematically accessible mass for the $U(1)_{B}$ benchmark.
We observe that for DUNE ND and SHiP, we can reach the Type-I seesaw band for some of our benchmark model choices. These limits extend well beyond the existing bounds for the HNL searches and are comparable to other recent proposals targeting the seesaw region (see e.g., Refs.~\cite{Abdullahi:2023gdj, Capozzi:2024pmh, Liu:2024fey, Wang:2024mrc, Drewes:2025ocf}). Note that in the minimal HNL scenario, that is, without additional interactions of HNLs beyond their mixing with active neutrinos, the HNL searches may not be able to reach the seesaw band, especially in the tau channel~\cite{Ballett:2019bgd, Berryman:2019dme, Feng:2024zfe}. 

In Fig.~\ref{fig:SensitivityZ}, we present the projected $Z'$ sensitivity reach at $90\%$ C.L.~in the $g_{B-L}$~–~$m_{Z'}$ (top left), $g_{B-3L_{\tau}}$~–~$m_{Z'}$ (top right), and $g_B$~–~$m_{Z'}$ (bottom) planes for the $N\to \nu\pi^0$ (solid) and $N\to \nu e^+e^-$ (dashed) channels. Overall, we observe consistent coverage between models because each of the models only differ by minor model-dependent factors. For the benchmark choice $m_{Z'}/m_N=2.1$ and $|U_{\tau}| = 10^{-3}$, which is safely below the existing experimental constraints, SBND achieves sensitivity of $90\%$ C.L.~around $g_X\sim 10^{-3}$ in the $\nu\pi^0$ mode for $U(1)_{B-L}$ in Fig.~\ref{fig:SensitivityZ}(a) and $U(1)_{B-3L_{\tau}}$ in Fig.~\ref{fig:SensitivityZ}(b), and in both channels for $U(1)_B$ in
Fig.~\ref{fig:SensitivityZ}(c). DarkQuest improves this reach, with the $\nu\pi^0$ channel probing couplings below $g_X\sim 10^{-3}$ in all three scenarios in Fig.~\ref{fig:SensitivityZ}. DUNE further extends the sensitivity, with the $\nu e^+e^-$ channel reaching below $g_X\sim 10^{-4}$ across the models. SHiP achieves the strongest reach, with sensitivity to $g_X\sim 5\times 10^{-6}$ for both $U(1)_{B-L}$ in Fig.~\ref{fig:SensitivityZ}(a) and $U(1)_{B-3L{\tau}}$ in 
Fig.~\ref{fig:SensitivityZ}(b), and down to $g_X\sim  10^{-6}$ for $U(1)_{B}$ 
in Fig.~\ref{fig:SensitivityZ}(c). DUNE ND and SHiP probe parameter space beyond current bounds for $U(1)_{B-L}$ in
Fig.~\ref{fig:SensitivityZ}(a). DarkQuest, DUNE ND, and SHiP do so for $U(1)_{B-3L{\tau}}$ in Fig.~\ref{fig:SensitivityZ}(b). All four experiments extend coverage beyond existing limits up to several orders of magnitude for $U(1)_B$ in
Fig.~\ref{fig:SensitivityZ}(c).

Before we summarize our results, a  few remarks are in order: 
\begin{itemize}
\item[$\bullet$] 
Thus far, we have only explored the clean decay channels $N\to \nu  e^+e^- ,\ \nu\pi^0$. Another significant channel could be $N\to \nu+$hadrons. Although its signature and kinematics may be similar to the $\pi^0\nu$ mode, some extra care may be needed for the background analysis, especially at the DUNE ND. Therefore, to be conservative, we will leave this and other potential contributions for future studies.
\item[$\bullet$]
Throughout our presentation, we have taken $N$ as a Majorana neutrino. In this case, $NN$ production through the $Z'$ decay is $p$-wave suppressed, i.e., exhibits the $\beta_N^3$ behavior as in Eq.~\eqref{eq:NN}. If we consider it as a (pseudo)Dirac neutrino instead, the corresponding production proceeds via an $s$-wave and scales as $\beta_N$, resulting in a larger production rate near the kinematic threshold $m_{Z'} \simeq 2 m_N$. Consequently, the signal sensitivity would be enhanced in the Dirac case.
\item[$\bullet$] 
It is important to note that, although we focus on the mixing $U_\tau$, our signals are flavor-independent since we do not require $\tau$ flavor identification, and the resultant bounds are thus largely applicable to $U_e$ and $U_\mu$ as well, up to the appropriate BR scaling. Only because the existing laboratory limits in the $U_e$ and $U_\mu$ sectors are much more stringent than in the $U_\tau$ sector~\cite{bolton, hostert}, our proposed HNL signal via the Drell-Yan production is not expected to buy us significantly new parameter space in the $U_e$ and $U_\mu$ sectors, and therefore, we decided not to show these sensitivities here. 
\item[$\bullet$] 
We add that our studies show  significant sensitivity to $Z'\to NN$, which is model-dependent as specified in the text. In all the $U(1)$ models we considered, this is indeed the dominant search channel, as the $Z'$ decay modes to SM final states are prompt and do not reach the detector (see Fig.~\ref{fig:GZp}).
In addition, for a suitable choice of the $U(1)_X$ charges, the $Z\to NN$ branching fraction can also be enhanced~\cite{Das:2017flq}, thus further increasing our sensitivity reach.
\end{itemize}

%%%%%%%%%%%%%%%%%%%\FloatBarrier
\section{Summary and Conclusions}
\label{sec:summary}

In this work, we demonstrated that Drell-Yan processes arising from deep inelastic scattering can play a significant role in searches for BSM physics at fixed-target or beam-dump experiments, where such processes have not been previously studied in the literature. 
Focusing on the current and future projects, we present our results in the context of ongoing experiments such as SBND, as well as future facilities including DarkQuest, DUNE and SHiP. 
As a case study, we explored the search sensitivity for HNLs  through the Drell-Yan resonance production of a light vector boson mediator: $pp\to Z'\to NN$. 

After setting up the models and theory parameters in Section \ref{Sec:Model}, we calculated the decay widths and the branching fractions for $Z'$ and $N$ to the observable final states (see  Figs.~\ref{fig:GZp} and \ref{fig:BR_UT_1}).  We commented on the existing bounds and our benchmark choice of the theory parameters.
We laid out the experimental settings and the detector specifications for the four benchmark experiments in Section \ref{sec:Benchmark}.
We performed detailed analyses for the signal and potential backgrounds in each experimental setting in Section \ref{sec:SB}.
Utilizing the on-shell resonant production, light vector bosons in the 2 GeV$-$20 GeV mass range are efficiently produced and promptly decay into pairs of HNLs. The energy spectra for $N$ are shown in Fig.~\ref{fig:mN_Event_Distributions}, which are typically harder than the SM background processes and thus serve as one of the main discriminators to separate out the signal. The resulting HNLs propagate to the detector, and subsequently decay into a variety of visible final states.
Among the visible final states, we singled out the channels $N\rightarrow\nu\pi^0$ and $N\rightarrow\nu e^+e^-$ 
as sensitive probes of the parameter space of $N$ for the leading $U_{\tau}$ mixing. 
Their characteristic energy distributions for DUNE ND were shown in Fig.~\ref{fig:mN_Event_Distributions_DUNE} for the three theoretical models, compared with the background expectations.

We presented our results for the sensitivity reach in Section \ref{Sensitivity}. Using the four experiments mentioned above and three theoretical benchmark models, we presented the sensitivities of the HNL signal at $90\%$ C.L. in the $U_\tau^2-m_N$ plane in Fig.~\ref{fig:Sensitivity_pinu}. Noting that the magnitude difference among models is primarily due to the benchmark coupling choices as in 
Eq.~(\ref{eq:coup}), we presented results for the fixed mass $m_{Z'} =  2.1$ GeV in Figs.~\ref{fig:Sensitivity_pinu}(a) and \ref{fig:Sensitivity_pinu}(b) for channels $N\to \nu\pi^0$ and $N\to \nu e^+e^-$, respectively. We also showed results for a fixed mass ratio $m_{Z'}/m_N = 2.1$ in Figs.~\ref{fig:Sensitivity_pinu}(c) and \ref{fig:Sensitivity_pinu}(d) for the same two decay channels. We find that the $\nu\pi^0$ channel is more sensitive than the $\nu e^+e^-$ channel for SBND, DarkQuest, and SHiP, whereas at DUNE ND, the $\nu e^+e^-$ channel is better. From the $\nu\pi^0$ channel in Figs.~\ref{fig:Sensitivity_pinu}(a) and \ref{fig:Sensitivity_pinu}(c), we found that SBND and DarkQuest can achieve sensitivity beyond the current neutrino mass-mixing limits, with both probing $|U_{\tau}| \gtrsim 10^{-3}~(3\times 10^{-4})$ for the $U(1)_{B-3L_{\tau}}$ $(U(1)_B)$ model benchmarks. We note in particular that sensitivity for SBND and DarkQuest can be achieved in the $U(1)_B$ model and similar models which do not face constraints from $e^+e^-$ colliders, allowing larger gauge couplings. From the $\nu e^+ e^-$ channel in Figs.~\ref{fig:Sensitivity_pinu}(b) and \ref{fig:Sensitivity_pinu}(d), we found that DUNE ND reaches the Type-I Seesaw band $|U_{\tau}|\sim 10^{-5}$ for the $U(1)_B$ benchmark. From Figs.~\ref{fig:Sensitivity_pinu}(a) and \ref{fig:Sensitivity_pinu}(c) for the $\nu \pi^0$ channel and Figs.~\ref{fig:Sensitivity_pinu}(b) and \ref{fig:Sensitivity_pinu}(d) for the $\nu e^+e^-$ channel, we found that SHiP can also reach the Type-I Seesaw bound for both $U(1)_{B-3L_{\tau}}$ and $U(1)_B$ benchmarks.

Finally, as shown in Fig.~\ref{fig:SensitivityZ}, for our chosen benchmark $|U_{\tau}| = 10^{-3}$ with a fixed mass ratio $m_{Z'}/m_N = 2.1$, and considering $U(1)_{B-L}$, $U(1)_{B-3L_{\tau}}$, and $U(1)_{B}$ parameter spaces, we found that both SBND and DarkQuest can probe gauge coupling $g_{X} \sim 10^{-3}$, DUNE ND can reach $g_{X} \sim 10^{-4}$, and SHiP can probe down to $g_{X}\sim 5\times 10^{-6}$ for all three models.

In conclusion, we found that at fixed-target and beam-dump experiments, the Drell-Yan production process can yield significantly more energetic final-state particles compared to the SM background, where neutrinos originate from less energetic meson decays. This enhanced energy allows us to probe the signal in a background-suppressed environment.
Our approach can be readily extended to light dark matter and other dark mediators within a broad class of light dark sector models. %

%

%

%%%%%%%%%%%%%%%%%%%%%%%%%%%%%%%%%%%
\section*{Acknowledgments}

We would like to thank Brian Batell, Vishvas Pandey, Maxim Pospelov,  Rohan Rajagopalan, and Zahra Tabrizi for helpful discussions. The work of TH and FB was supported in part by the US Department of Energy under grant No. DE-SC0007914 and in part by Pitt PACC. The work of PSBD was supported in part
by the US Department of Energy under grant No. DE-SC0017987. This work of BD and AK is supported by the U.S. Department of Energy
Grant~DE-SC0010813.

\FloatBarrier

\appendix
\section{HNL Decay Rates} 
\label{app: HNL Decays}

HNL decays are mediated by the mixing angle $U_{\alpha}$ (where $\alpha = e, \mu, \tau$) with the SM neutrinos, which induce both CC and NC interactions. The final states can be either leptonic or hadronic. In the latter case, for $m_N\lesssim 1$ GeV, the decays are into single-meson states, whereas for $m_N\gtrsim 1$ GeV, the decays into multi-meson states become relevant. For details, see e.g., Refs.~\cite{Gorbunov:2007ak, Bondarenko:2018ptm, Ballett:2019bgd, Coloma:2020lgy, Capozzi:2024pmh, Feng:2024zfe, Atre:2009rg, Helo:2010cw}. Here we have assumed HNLs to be Majorana particles and have properly taken into account both lepton-number-conserving and violating decay modes.

\subsubsection*{Leptonic Channels:}
\begin{align}
\Gamma(N \to \nu_\beta \bar{\nu}_\beta \nu_{\alpha}) &= \Gamma(N \to \nu_\beta \bar{\nu}_\beta\bar{\nu}_{\alpha}) =  
\frac{G_F^2m_N^5}{192\pi^3}|U_{\alpha}|^2 \, ,\\
\Gamma(N\to \nu_{\alpha}\gamma) &= \Gamma(N\to \bar{\nu}_{\alpha}\gamma) = 
\frac{9\alpha_{\rm em}G_F^2m_N^5}{512\pi^4}|U_{\alpha}|^2 \, ,\\
\Gamma(N \to \nu_{\alpha}l^+_{\beta}l_{\beta}^-) &= \Gamma(N \to \bar{\nu}_{\alpha}l^+_{\beta}l_{\beta}^-) = \frac{G_F^2m_N^5}{192\pi^3}|U_{\alpha}|^2\left[(C_1+2s_W^2\delta_{\alpha\beta})f_1(x_\beta)\nonumber \right.\\
& \hspace{10em} \left. +(C_2+s_W^2\delta_{\alpha\beta})f_2(x_\beta)  \right]
\, , \\
\Gamma(N \to l^{\mp}_{\alpha \neq \beta}l_{\beta}^{\pm}\nu) &= \Gamma(N \to l^{\mp}_{\alpha \neq\beta}l_{\beta}^{\pm}\bar{\nu}) = 
\frac{G_F^2m_N^5}{192\pi^3}\left[|U_\alpha|^2I_1(x_\alpha,0,x_\beta)+|U_\beta|^2I_1(x_\beta, 0,x_\alpha) \right] \, ,
\end{align}
where $x_\alpha=m_{\alpha}/m_N$, $G_F$ is the Fermi constant, and 
\begin{align}
   C_1=\frac{1}{4}(1-4s_W^2+8s_W^4), \quad   C_2=\frac{1}{2}s_W^2(2s_W^2-1) \, ,
\end{align}
with $s_W^2\equiv \sin^2\theta_w\simeq 0.2312$~\cite{ParticleDataGroup:2024cfk} being the weak mixing angle. The kinematic functions are given by~\cite{Atre:2009rg}
\begin{align}
    I_1(x, y, z) &= 12\int_{(x + y)^2}^{(1 - z)^2}\frac{ds}{s}(s - x^2 - y^2)(1 + z^2 - s)\lambda^{1/2}(s, x^2, y^2)\lambda^{1/2}(1, s, z^2) \, ,\\
    f_1(x) & = (1-14x^2-2x^4-12x^6)\sqrt{1-4x^2}+12x^4(x^4-1)L(x)\, , \\
    f_2(x) & = 4\left[x^2(2+10x^2-12x^4)\sqrt{1-4x^2}+6x^4(1-2x^2+2x^4)L(x) \right] \, , \\
    {\rm where}\quad 
    L(x) &= \log{\left[\frac{1 - 3x^2 - (1 - x^2)\sqrt{1 - 4x^2}}{x^2(1 + \sqrt{1 - 4x^2})}\right]} \, , \\
    {\rm and}\quad \lambda(x, y,z) & = x^2 + y^2 +z^2- 2xy - 2yz - 2xz \, .
\end{align} 

\subsubsection*{Single-Meson Channels:} 
\begin{align}
\Gamma(N \to P^0\nu_{\alpha}) &= \Gamma(N \to P^0\bar{\nu}_{\alpha})=\frac{G_F^2m_N^3}{32\pi}f_{P}^2|U_{\alpha}|^2(1 - x_{P}^2)^2 \, , \\
\Gamma(N \to P^{\pm}l_{\alpha}^{\mp}) &=  \frac{G_F^2m_N^3}{16\pi}f_P^2|U_{\alpha}|^2|V_{qq'}|^2\lambda^{\frac{1}{2}}(1, x_P^2, x_\alpha^2)\left[1 - x_P^2 - x_\alpha^2(2 + x_P^2 - x_\alpha^2)\right]\, , \\
\Gamma(N \to V^0\nu_{\alpha}) &= \Gamma(N \to V^0\bar{\nu}_{\alpha})=\frac{G_F^2m_N^3}{32\pi}f_V^2|U_{\alpha}|^2\kappa_V^2(1 + 2x_V^2)(1 - x_V^2)^2 \, , \\
\Gamma(N \to V^{\pm}l_\alpha^{\mp}) &= \frac{G_F^2m_N^3}{16\pi}f_V^2|U_{\alpha}|^2|V_{qq'}|^2\lambda^{\frac{1}{2}}(1, x_V^2, x_\alpha^2)[(1 - x_V^2)(1 + 2x_V^2) + x_\alpha^2(x_V^2 + x_\alpha^2 - 2)]\, , 
\end{align}
where $x_{P,V}=m_{P,V}/m_N$, and $V_{qq'}$ is the relevant CKM matrix element. Here, $P^0=\pi^0, \eta,\eta'$ are the neutral and $P^\pm=\pi^\pm,K^\pm, D^\pm, D_s^\pm,B^\pm,B_c^\pm$ the charged pseudoscalar mesons, and   $V^0=\rho,\omega,\phi,J/\psi$ are the neutral and $V^\pm=\rho^\pm,K^{*\pm},D^{*\pm},D_s^{*\pm}$ are the charged vector mesons relevant for the mass range shown in Fig.~\ref{fig:BR_UT_1}. We have used the PDG average for their mass values~\cite{ParticleDataGroup:2024cfk}.  For the neutral vector meson couplings,
\begin{align}
   \kappa_\rho=1-2s_W^2, \quad \kappa_\omega=-\frac{2s_W^2}{3}, \quad  \kappa_\phi=-\sqrt 2\left(\frac{1}{2}-\frac{2s_W^2}{3} \right), \quad \kappa_{J/\psi}=1-\frac{8}{3}s_W^2 .
\end{align}
The meson decay constants are given in Table~\ref{tab:decay} where we use the experimental average whenever available, and theory predictions otherwise. 
\begin{table}[h!]
\centering
\begin{tabular}{|c|c|} \hline
$P$ & $f_P$ (MeV) \\ \hline
$\pi^0, \pi^\pm$ & 130.5~\cite{ParticleDataGroup:2024cfk} \\ \hline
$\eta$ & 81.6~\cite{Coloma:2020lgy} \\ \hline
$\eta'$ & $-94.6$~\cite{Coloma:2020lgy} \\ \hline
$\eta^c$ & $387$~\cite{Becirevic:2013bsa} \\ \hline
$K^\pm$ & 155.7~\cite{ParticleDataGroup:2024cfk}  \\ \hline
$D^{\pm}$ & 210.4~\cite{FlavourLatticeAveragingGroupFLAG:2024oxs} \\ \hline
$D_s^{\pm}$ & 247.7~\cite{FlavourLatticeAveragingGroupFLAG:2024oxs} \\ \hline
$B^\pm$ & 190~\cite{FlavourLatticeAveragingGroupFLAG:2024oxs} \\  \hline
$B_c^\pm$ & 434~\cite{Colquhoun:2015oha} \\  \hline
\end{tabular}
\hspace{3em}
\begin{tabular}{|c|c|}
\hline
$V$ & $f_V$ (MeV) \\ \hline
$\rho^0, \rho^\pm$ & 210~\cite{Bharucha:2015bzk} \\ \hline
$\omega$ & 197~\cite{Bharucha:2015bzk} \\ \hline
$\phi$ & 229~\cite{Chakraborty:2017hry}  \\ \hline
$J/\psi$ & 418~\cite{Becirevic:2013bsa} \\ \hline
$K^{*\pm}$ & 204~\cite{Bharucha:2015bzk} \\ \hline
$D^{*\pm}$ & 242~\cite{Gelhausen:2013wia} \\ \hline
$D_s^{*\pm}$ & 293~\cite{Gelhausen:2013wia} \\ \hline
\end{tabular}
\caption{The pseudoscalar (left table) and vector (right table) meson decay constants used in our analysis.}
\label{tab:decay}
\end{table}

\subsubsection*{Multi-Meson Channels:} 
HNLs with $m_N \gtrsim 1$ GeV can decay into quarks which then hadronize into multiple mesons. The tree-level NC and CC decays into quarks are respectively given by 
\begin{align}
    \Gamma(N\to \nu_\alpha q\bar{q}) &= \Gamma(N\to \bar{\nu}_\alpha q\bar{q}) = \frac{G_F^2m_N^5}{64\pi^3}|U_\alpha|^2\left[C_1^qf_1(x_\alpha)+C_2^q f_2(x_\alpha) \right] \, , \\
    \Gamma(N\to l^\pm_\alpha q \bar{q}') &= \frac{G_F^2m_N^5}{64\pi^3}|U_\alpha|^2 |V_{qq'}|^2 I_1(x_\alpha,x_q,x_{q'}) \, ,
\end{align}
where the coefficients $C_{1,2}^q$ are given by 
\begin{align}
    C_1^{u,c} = & \frac{1}{4}\left(1-\frac{8}{3}s_W^2+\frac{32}{9}s_W^4 \right) \, , \quad  C_2^{u,c} =  \frac{1}{3}s_W^2\left(\frac{4}{3}s_W^2-1 \right) \, , \\
     C_1^{d,s,b} = & \frac{1}{4}\left(1-\frac{4}{3}s_W^2+\frac{8}{9}s_W^4 \right) \, , \quad  C_2^{d,s,b} =  \frac{1}{6}s_W^2\left(\frac{2}{3}s_W^2-1 \right) \, .
\end{align}
The total hadronic decay width can be estimated using these tree-level decay modes along with a
QCD loop correction to account for the hadronization process~\cite{Bondarenko:2018ptm, Coloma:2020lgy}. The loop correction is estimated from hadronic tau decays, and is defined as 
\begin{align}
    1+\Delta_{\rm QCD} \equiv \frac{\Gamma(\tau\to \nu_\tau+{\rm hadrons})}{\Gamma_{\rm tree}(\tau\to \nu_\tau u\bar{d})+\Gamma_{\rm tree}(\tau\to \nu_\tau u \bar{s})} \, , 
\end{align}
where the correction up to ${\cal O}(\alpha_s^3)$ has been calculated~\cite{Gorishnii:1990vf}: 
\begin{align}
    \Delta_{\rm QCD} = \frac{\alpha_s}{\pi}+5.2\frac{\alpha_s^2}{\pi^2}+26.4\frac{\alpha_s^3}{\pi^3} \, ,
\end{align}
where $\alpha_s\equiv g_s^2/4\pi$, with $g_s$ being the strong coupling. The running of $\alpha_s(m_N)$ is taken into account~\cite{ParticleDataGroup:2024cfk}. For the $\nu s\bar{s}$ channel, an additional suppression factor $\sqrt{1-4m_K^2/m_N^2}$ is included to not overestimate its contribution for $m_N<2m_K$. The same approach is also applied to the $\tau u\bar{d}$ and $\tau u\bar{s}$ final states, with kinematic thresholds at $m_\tau+2m_\pi$ and $m_\tau+m_\pi+m_K$, respectively~\cite{Feng:2024zfe}.

Apart from the SM decay modes discussed above, HNLs have additional decay modes mediated via the $Z'$ boson. We have checked that these additional decay modes  become relevant only for large enough gauge coupling values $g_X\gtrsim 10^{-2}$ (allowed e.g., in the $U(1)_B$ model) and small enough $Z'$ masses $m_{Z'}\lesssim 3$ GeV. This gives an ${\cal O}(1)$ correction to the HNL branching fractions shown in Fig.~\ref{fig:BR_UT_1} and has a negligible effect on the  sensitivities shown in Figs.~\ref{fig:Sensitivity_pinu} and \ref{fig:SensitivityZ}.     
\FloatBarrier

\FloatBarrier
\bibliographystyle{JHEP}
\bibliography{references}

\end{document}